\documentclass[aps,preprint,epsfig,rotate,showpacs]{revtex4}
\usepackage{epsfig}
\usepackage{graphicx}
\usepackage{amsmath}
\usepackage{graphicx}
\usepackage{epstopdf}

\newcommand{\beq}{\begin{eqnarray}}
\newcommand{\eeq}{\end{eqnarray}}
\begin{document}
\title{Transient Quantum Coherent Response to a Partially Coherent Radiation Field}
\author{Z. S. Sadeq and P. Brumer*}
\affiliation{Chemical Physics Theory Group, Department of Chemistry \\ and
Center for Quantum Information and Quantum Control\\
University of Toronto, Toronto, M5S 3H6, Canada}
\date{\today}
\begin{abstract}
The response of an arbitrary closed quantum system  to a partially coherent electric field
 is investigated, with a focus on the transient coherences in the system.
As a model we examine, both perturbatively and numerically, the coherences induced in a three level $V$ system.   Both rapid turn-on and pulsed turn-on effects are investigated.  The effect of a long and incoherent
  pulse is also considered, demonstrating that during the pulse the system shows a coherent response which
   reduces after the pulse is over. Both the pulsed scenario and the thermally broadened CW case approach
    a mixed state in the long time limit, with rates dictated by the adjacent level spacings and the coherence time of the light,
    and via a mechanism that is distinctly difference from traditional decoherence.
     These two excitation scenarios are also explored for a minimal ``toy" model of
      the electronic levels in pigment protein complex PC645 by both a collisionally broadened CW laser and by a noisy pulse, where unexpectedly
      long transient coherence times are observed and explained.  The significance of environmentally induced decoherence is noted.

\end{abstract}
\email{pbrumer@chem.utoronto.ca}
\maketitle

\section{Introduction}

Recent advances \cite{col,engel} in coherent femtosecond nonlinear spectroscopy have shown evidence
of long-lived  coherences in photosynthetic systems. However, in nature biological systems
are exposed  to incoherent sources of light, such as sunlight, as opposed
to coherent femtosecond sources. Therefore, an understanding of excitation under incoherent conditions is crucial. In such cases,  the \textit{long time} result of incoherent irradiation, using both quantum and classical light,
leads to a mixed state of the system \cite{brumer,onephoton}, a result  discussed further in the
literature in the context of photosynthetic light harvesting complexes \cite{mancal,leo,Han}.  However, work has been done, for example,  on incoherent or noisy perturbations in quantum
optics where some have found that incoherent or noisy electric fields can 
generate transient coherences in three 
level atoms \cite{aharony, kozlov, scully}.  
Hence, understanding the approach to the mixed state, and 
its dependence on system parameters is important, and is 
the subject of this paper. 

Here, transient excited state coherences manifest upon
the perturbative sudden turn-on irradiation of an isolated quantum system are
discussed and examined in a V level system.  These coherences are
shown to  become negligible in the long time limit relative to the
populations so that the system approaches, in this way, a mixed
state. The dynamics of long and noisy pulses incident on generic
systems is also investigated. These pulses induce coherences on a
time scale dependent on the excited state period and the pulse
duration. However, after the pulse is over these coherences become
damped and  irrelevant relative to the magnitude of populations.
In both cases, the magnitude of the spacing between the excited
levels is shown to play an important role. Further, in both cases,
the approach to the mixed state is dramatically different than that
associated with decoherence experienced by a prepared superposition state
that is in contact with a thermal bath (see, e.g., \cite{Elran1,Elran2}).

We consider the response of a system to two sample sources
of incoherent light.  The first  models  incoherent light
corresponding to a collisionally broadened CW source \cite{loudon}
that is turned on abruptly. This incoherent light source is
characterized by a two time correlation function \cite{brumer}:
\begin{equation}
\langle\varepsilon(t')\varepsilon^{*}(t'')\rangle=\varepsilon_{0}^{2}e^{-i\omega_{0}\left(t'-t''\right)}e^{-\frac{|t'-t''|}{\tau_{d}}}
\label{corr1}
\end{equation}
where $\omega_{0}$ is the frequency center of the radiation, and $\varepsilon_{0}^{2}$ is the field intensity. The coherence time of this
radiation is given by $\tau_{d}=\hbar/kT$
where $k$ is Boltzmann's constant and $T$ is temperature \cite{wolf, mandelwolf}.
Hence, a room temperature source at  $T=300\text{ K}$ gives  $\tau_{d}\simeq 25$
fs whereas a  source with $T=5800$
K, gives $\tau_{d}\simeq1.32$ fs. This corresponds to a classical model of incoherent light; the full quantum version has qualitatively similar behavior to the correlation function in Eq. (\ref{corr1})  at these temperatures \cite{loudon,wolf,mandelwolf}.

A second source examined below is a noisy
pulse,  a Gaussian pulse with a phase jitter designed to model an incoherent
light pulse, given by \cite{brumerbook}:

\begin{equation}
\langle\varepsilon(t')\varepsilon^{*}(t'')\rangle=\varepsilon_{0}^{2}e^{-\frac{\left(t'-t_{m}\right)^{2}}{\tau_{p}^{2}}}e^{-\frac{\left(t''-t_{m}\right)^{2}}{\tau_{p}^{2}}}e^{i\omega_{0}\left(t''-t'\right)}e^{-\frac{\left(t'-t''\right)^{2}}{2\tau_{d}^{2}}}
\label{corr2}
\end{equation}

Natural processes experiencing incoherent light  can be modeled
using Eq. (\ref{corr2}) with the pulse duration $\tau_{p}$  being
much larger than the coherence parameter, $\tau_{d}$. In this
limit Eq. (\ref{corr2}) behaves like Eq. (\ref{corr1}), but with a
smooth turn-on of the field. The attractiveness of using a long
incoherent pulse is two fold: (1) it has no sudden ``turn-on
effect'' that leads to artificial initial coherence, and (2)  it
offers a reasonable long time result which models radiation
induced decoherence \cite{brumerbook}.  In addition, in the pulsed
scenario, there are three time scales of interest: $\tau_{d}$ the
pulse coherence time, $\tau_{p}$ the pulse duration, and
$\tau_{c}$ the characteristic time scale of the system, usually
equal to the inverse of the level spacing.  In the simulations
below we assume that the pulse duration $\tau_p$ is much longer
than both $\tau_d$ and $\tau_c$. In addition since, 
in a natural environment, the system is initially in a 
stationary mixed state, we consider the light to be incident
on a single molecular eigenstate, i.e. a representative element of
the ensemble.    Finally,  note that both Eqs.
(\ref{corr1}) and (\ref{corr2}) are broadened about a single
frequency $\omega_{0}$. Typical natural light, however, is a
mixture of different $\omega_{0}$ values.

Below we consider both analytic first
order perturbation theory results as well as numerical results for
these cases.  In the numerical implementation, the correlation
functions    [Eqs. (\ref{corr1}) and (\ref{corr2})]   are produced
as averages over many individual realizations built from a Wiener
process (see Appendix A), i.e., electric fields $\varepsilon(t)$ with suitable
distribution of jumps in phase \cite{gardiner}.   Specifically,
the collisionally broadened CW source is reproduced by perturbing
a classical oscillator's frequency with Wiener noise. The noisy
pulse is generated in a similar fashion, except that a Gaussian
envelope is placed over the noisy oscillator.

A remark is in order with respect to this approach. Physically,
the meaningful quantity is the molecular density matrix  that
arises from the effect of the ensemble of electric fields that
satisfy the statistics in  Eqs. (\ref{corr1}) and (\ref{corr2}).
Such electric field statistics can arise from a wide variety of
different types of realizations.  For example, it has been
suggested that coherent fs pulses may be a meaningful basis from
which to build such realizations \cite{fleming}.
 To do so  requires that the pulses be added together \textit{incoherently} casting doubt  on the relevance of the coherence observed in
  any single realization.    In addition, such a fs basis, unlike our choice of phase interruptions to a CW source, has no physical justification.

The paper is organized as follows.   Section II  considers
the short time response of a closed quantum system, both perturbatively
and numerically, to a collisionally broadened CW source, Eq. (\ref{corr1}) after sudden turn-on.  Section III
deals with the case of long incoherent pulses, Eq. (\ref{corr2}) incident on  a model molecular
system, both perturbatively and numerically.  A toy model of the electronic energy levels of the pigment-protein complex PC645 is used in both sections as an interesting example. The paper is summarized in Sec. IV.

\section{Short Time Response of an Energy Level to Incoherent Light}

The effect of CW incoherent light incident on a molecule in an energy eigenstate and
starting  in the infinite past, is treated in detail in Refs.
\cite{brumer,onephoton}.  There, the effect was to produce, in the long time limit,  a
mixture of stationary system energy eigenstates.  Absent from this
treatment was the dynamics by which this came about, which is the
subject of this section.  Note that given the weakness of natural
light (such as sunlight) a first-order perturbative treatment is
applicable.

\subsection{Perturbative Treatment (Analytic)}

The total Hamiltonian of the system and radiation is  given by $H
= H_{0} - \mu\varepsilon(t)$, where $\mu$ is the dipole operator
and $\varepsilon(t)$ is the electric field. Using the standard
dipole approximation  and treating the light-molecule interaction
to first order gives the following expression \cite{brumer} for
the excited state density matrix:

\begin{widetext}
\begin{equation}
\rho_{e}=\sum_{i,j}c_{i}c_{j}^{*}|i\rangle\langle j|e^{-i\omega_{ij}t}\int_{t_{0}}^{t}dt'e^{i\omega_{ig}t'}\int_{t_{0}}^{t}dt''e^{-i\omega_{jg}t''}\langle\varepsilon(t')\varepsilon^{*}(t'')\rangle.
\label{den1}
\end{equation}
\end{widetext}
Here, the time independent coefficients $c_{i},c_{j}^{*}$ are
given by  $c_{i} =\varepsilon_{0}\langle i|\mu|g\rangle/i\hbar$
where the  $|i\rangle$ are the energy eigenstates of the system at
energy $ E_i$ and  $\omega_{jg}=(E_{j}-E_{g})/\hbar$ is the
frequency difference between the $j$th state and the ground state.
The initial condition has the system in the ground state, i.e.,
$\rho(0)= |g\rangle\langle g|$. Equation (\ref{den1}) has been
obtained assuming that the ensemble average over the product of
the electric fields and the time integrals commute. The validity
of this assumption is confirmed numerically in Sec. IIB.

 For the collisionally broadened CW source, we insert Eq. (\ref{corr1})
  for $\langle\varepsilon(t')\varepsilon^{*}(t'')\rangle$ into
 Eq. (\ref{den1})
and compute the two time integrals.  The integral can be written
as  four distinct terms, so that the excited state density matrix
is:

\begin{equation}
\rho_{e}=\sum_{i,j}c_{i}c_{j}^{*}|E_{i}\rangle\langle E_{j}|\left(\eta_{LT}^{ij}+\eta_{1}(ij)+\eta_{2}(ij)+\eta_{3}(ij)\right) \label{eq6}
\end{equation}

\begin{equation}
\eta_{LT}^{ij}=i\mathcal{U}\left(\omega_{jg}\right)\frac{e^{-i\omega_{ij}t}-1}{\omega_{ij}}
\end{equation}

\begin{equation}
\eta_{1}(ij)=-\frac{\mathcal{U}\left(\omega_{jg}\right)}{2}\frac{e^{t\left(i\delta_{j}-\frac{1}{\tau_{d}}\right)}-e^{-i\omega_{ij}t}}{i\delta_{i}-\frac{1}{\tau_{d}}}
\label{eta1}
\end{equation}

\begin{equation}
\eta_{2}(ij)=-\frac{\mathcal{U}\left(\omega_{jg}\right)}{2}\left(\frac{1-e^{-i(\delta_{j}+\omega_{ij})t}e^{-\frac{t}{\tau_{d}}}}{\frac{1}{\tau_{d}}+i\delta_{i}}\right)
\label{eta2}
\end{equation}

\begin{equation}
\eta_{3}(ij)=iR\left(\delta_{j}\right)\left(\frac{1-e^{-i(\delta_{j}+\omega_{ij})t}e^{-\frac{t}{\tau_{d}}}}{\frac{1}{\tau_{d}}+i\delta_{i}}-\frac{e^{i\delta_{j}t}e^{-\frac{t}{\tau_{d}}}-e^{-i\omega_{ij}t}}{i\delta_{i}-\frac{1}{\tau_{d}}}\right)
\label{eta3}
\end{equation}
For simplicity, we have set $t_{0} = 0$; averaging over $t_0$ is
discussed in Appendix B. Here, $\delta_{k}=\omega_{kg}-\omega_{0}$ gives the
detuning of the $|g\rangle \rightarrow |k\rangle$ transition
frequency from the central laser frequency,
 and $R\left(\delta_{j}\right)= \delta_{j}/(\frac{1}{\tau_{d}^{2}}+\delta_{j}^{2})$ is a  Lorentzian that
  is dependent on the detuning   and on the coherence parameter $\tau_d$.

The first term,  $\eta_{LT}^{ij}$, tends to dominate at long times
and the $\eta_{k}(ij)$  contain transient terms that decay  on the field coherence time, $\tau_{d}$.
The terms $\eta_{LT}^{ij},\eta_{1},\eta_{2}$ are  preceded by a
Lorentzian $\mathcal{U}\left(\omega\right)$ that is characteristic of the Wiener process underlying the electric field statistics:
\begin{equation}
\mathcal{U}\left(\omega\right)=\frac{2\tau_{d}}{1+\tau_{d}^{2}\left(\omega-\omega_{0}\right)^{2}}
\label{mathcalU}
\end{equation}
In the limit of $\tau_{d} \rightarrow \infty$,
$\mathcal{U}(\omega)$ approaches $\delta(\omega-\omega_{0})$, thus
converging to the Fourier transform of a conventional CW laser.

As shown below, $\eta_{LT}^{ii}$ is a term that contributes significantly to the populations as time increases.  Specifically, in the limit of $\omega_{ij}\rightarrow0$,
\begin{equation}
\eta_{LT}^{ii}=\mathcal{U}\left(\omega_{ig}\right)t
\label{etaLT}
\end{equation}
Alternatively, this result can be obtained by solving Eq. (\ref{den1}) in
the limit of $j \rightarrow i$.

By examining Eqs. (\ref{eq6}) - (\ref{eta3})  after a time,
$t\gg\tau_{d}$, one sees that the contributions from the
$\eta_{k}, k=1,2,3$ are either negligible or  oscillate on a time
scale of the inverse of the level spacing, similar to the
$\eta_{LT}^{ij}$ term.  As the system evolves, the populations
grow at a constant rate, while the amplitude of the coherence
become fixed (but non-zero) after time $\tau_{d}$.    This implies that at some long time ($t\gg
\tau_{d}$), the populations become extremely large relative to
coherences, i.e. it is in this way that the system reaches a mixed
state. It also indicates that at long times the off-diagonal
elements $\rho_{ij}$ of $\rho_{e}$ are usually  small but non-zero.   Indeed, since the magnitude of $\eta^{i_j}_{LT}$ is inversely
proportional to $\omega_{ij}$,  nearby levels can display
large coherences. These
off-diagonal elements are a direct reflection of the open nature
of the quantum system; that the system is coupled to the radiative
bath, resulting in a system-bath coupling that leads to a
non-diagonal system density matrix  in the original system energy
basis \cite{leo}. This is an example of
``canonical-nontypicality'' and is of particular interest in
topics like one-photon phase control \cite{pachon-brumer}

The expression for populations is in agreement with that in Ref.
\cite{mancal}. However, they are in contrast with our earlier results of
Ref. \cite{brumer} that give the density matrix as diagonal in the
energy eigenbasis.  The difference results from the  choice,
in Ref. \cite{brumer}, of a turn-on in the infinite past and of a
continuous energy distribution.  Nonetheless, the qualitative
result here  is the same as that in Ref.  \cite{brumer}, i.e., the
system is found in a mixed state at long times.

Issues of the introduction of a specific laser turn-on time are
discussed in Appendix B where averaging over turn-on time $t_{0}$
is shown not to  eliminate the small coherences encountered
in this approach. In addition, Appendix B discusses conditions
under which the long time $\rho_{ij}$ are maximal (but still
vanishingly small compared to the populations).

These results indicate (1) the nature of the closed
system transient coherent response to the sudden turn-on of an
incoherent field, (2) the linearity of the population growth due to the
diffusion process underlying the radiation field, (3) that the
coherences so induced do survive as time evolves, but (4) that populations become much larger than the  coherences,
so that the system effectively  approaches a mixed state.

Sample computations quantifying these results, and allowing for an
analysis of the dependence on system properties,  are provided
below.

\subsection{ Perturbative Treatment (Numerical)}

Although the perturbation associated with natural sunlight is
weak, and hence Eqs.  (\ref{eq6}) - (\ref{eta3})   are expected to
give virtually exact results, numerical studies are necessary for
the pulsed case considered below (Sect. III) and for confirmation
of the validity of ordering of the time integral and  ensemble averaging
assumed above.  Below, the perturbation theory results are
compared to numerical results obtained by solving the von Neumann
equation of a $V$ level system:

\begin{equation}
\frac{d\rho}{dt}=\frac{i}{\hbar}\left[\rho,H\right]
\end{equation}
where the Hamiltonian  is given by:

\begin{equation}
H=\left(\begin{array}{ccc}
\omega_{g} & -\mu \varepsilon(t) & -\mu \varepsilon(t)\\
-\mu\varepsilon^{*}(t) & \omega_{1} & 0\\
-\mu\varepsilon^{*}(t) & 0 & \omega_{2}
\end{array}\right)
\end{equation}
and where the star denotes the complex conjugate.

We generate collisionally broadened CW sources by producing a set
of fields   $\{\varepsilon(t)\}$, which obeys the statistics  in
Eq. (\ref{corr1}). This was done by  changing the phase at random
times with the interruption times selected  from a Wiener
distribution, and  phase changes  chosen from a uniform
distribution. Details of this algorithm are in Appendix A.

\begin{figure}[h]
\begin{center}
\includegraphics[scale=0.3]{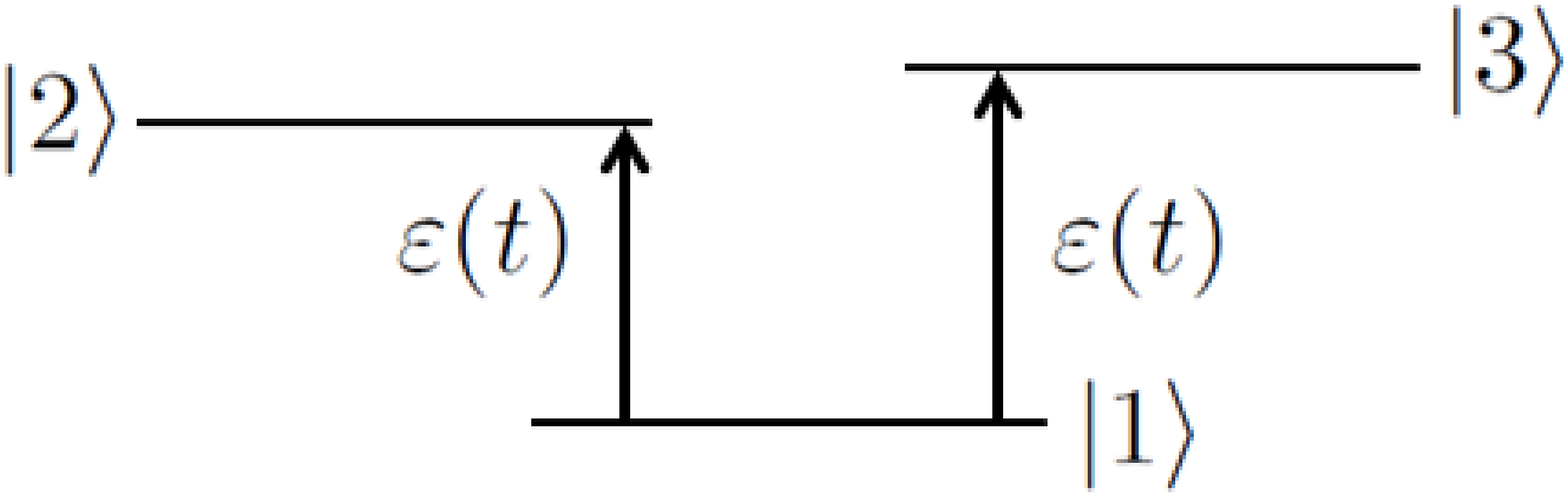}
\caption{Three level system used in our numerical scheme. Population
is initially in the ground state $|1\rangle$ and is pumped to
two non-degenerate excited states, $|2\rangle$ and $|3\rangle$.  The frequency difference between
the two excited states $\omega_{32}$ determines a characteristic
excited state timescale $\tau_{c}=2\pi/\omega_{32}$.}
\label{pprfig1}
\end{center}
\end{figure}

A model $V$ level system, illustrated in Fig.  \ref{pprfig1}, with
initial population in the ground state was subjected to a set of
fields  $\{\varepsilon(t)\}$. An average over the set of
individual realizations of $\varepsilon(t)$ gives the resultant
$\rho$.  To compare to the perturbative result obtained in Sec.
IIA we focus on the excited populations
 $\rho_{22},\rho_{33}$, and  coherences
between excited states  $\rho_{23}$.

\begin{figure}[h]
\begin{center}
\includegraphics[scale=0.4]{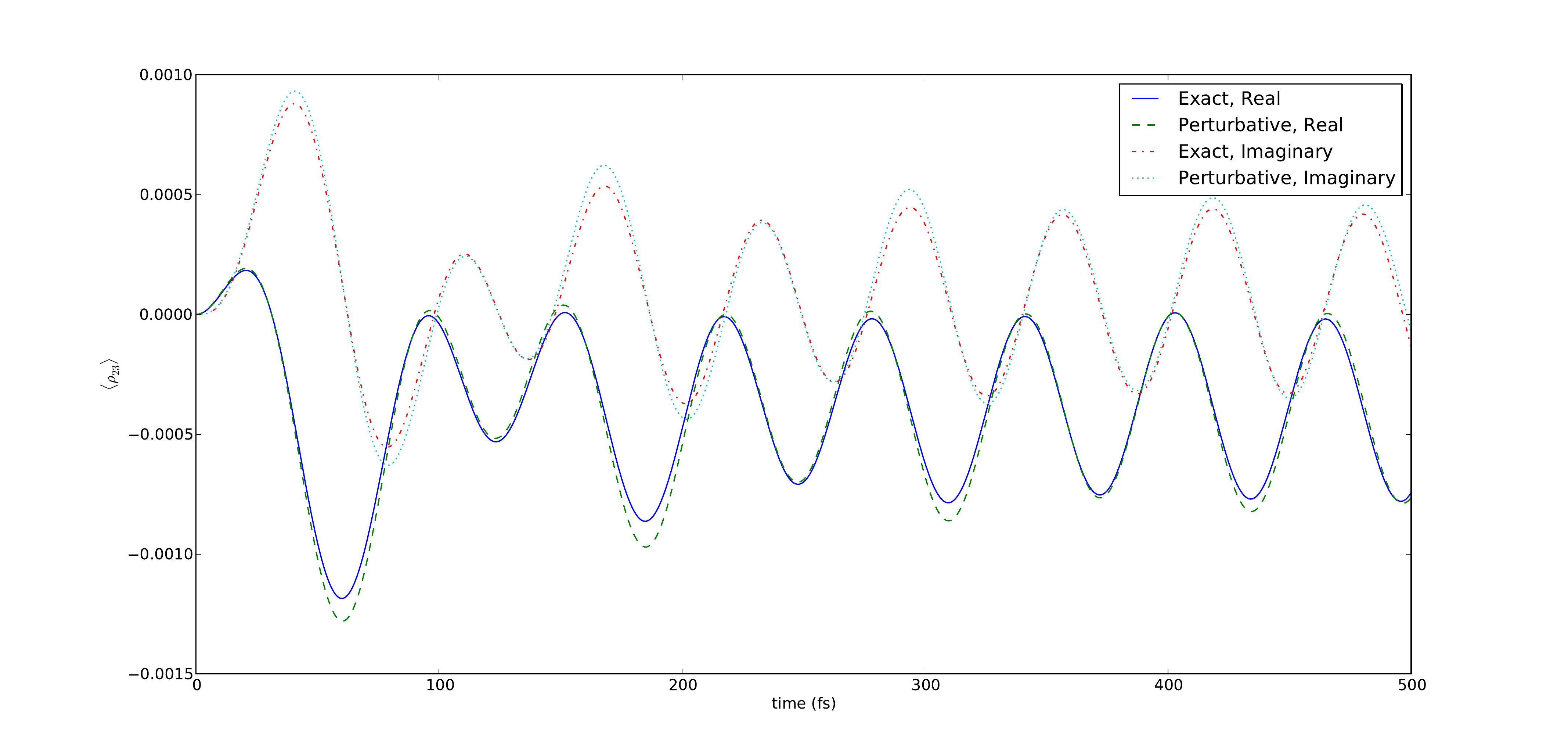} \includegraphics[scale=0.5]{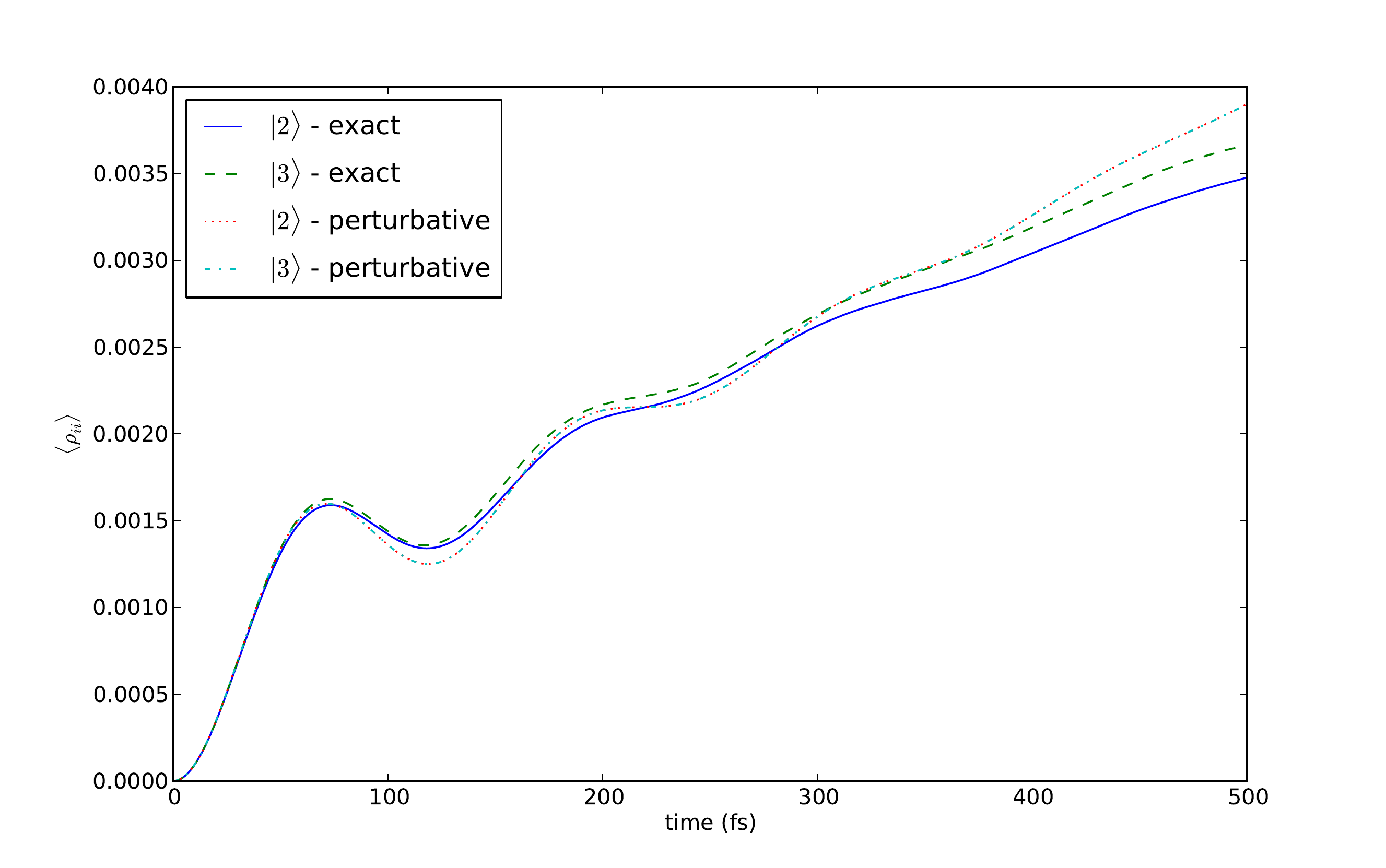}
\caption{(Top) Ensemble averaged excited state coherence
($\langle\rho_{23}\rangle$) plotted versus perturbative
coherences for a three level ladder system excited by thermally broadened CW source. (Bottom) Excited state populations for both
perturbative and numerical results for a three level ladder system excited by thermally broadened CW source. $\tau_{c} = 60$ fs and
$\tau_{d}=120$ fs for both figures.}
\label{popcohcomp}
\end{center}
\end{figure}

Figure \ref{popcohcomp} shows the perturbative vs. numerical
calculations for  the excited state populations and coherences for
a fixed value of  $\tau_{d} = 120$ fs.  This $\tau_d$ value, which
is two orders of magnitude larger than that of solar radiation, is
examined for convenience only.    The frequency of radiation used
is chosen to be $\omega_{0} = (\omega_{31}+\omega_{21})/2$ so as
to excite both transitions equally and here and below, unless otherwise indicated, $\mu\varepsilon_0/\hbar =$1THz.      Transient field
induced terms are clear at short times, but after $t\gg\tau_{d}$
the system oscillates at the level spacing, here  chosen to be
$\tau_{c} = \frac{2\pi}{\omega_{32}}\simeq60$ fs. The ensemble
averaged excited state populations show  linear growth, agreeing
with the dynamics predicted by the perturbative result derived
earlier. The perturbative and numerical  solutions show excellent
agreement for coherences while the numerical solution has a
slightly smaller slope to that of the perturbative result for
populations. This is due to the inability to generate the exact
correlation function accurately, as discussed in Appendix A.

\begin{figure}[h]
\begin{center}
\includegraphics[scale=0.39]{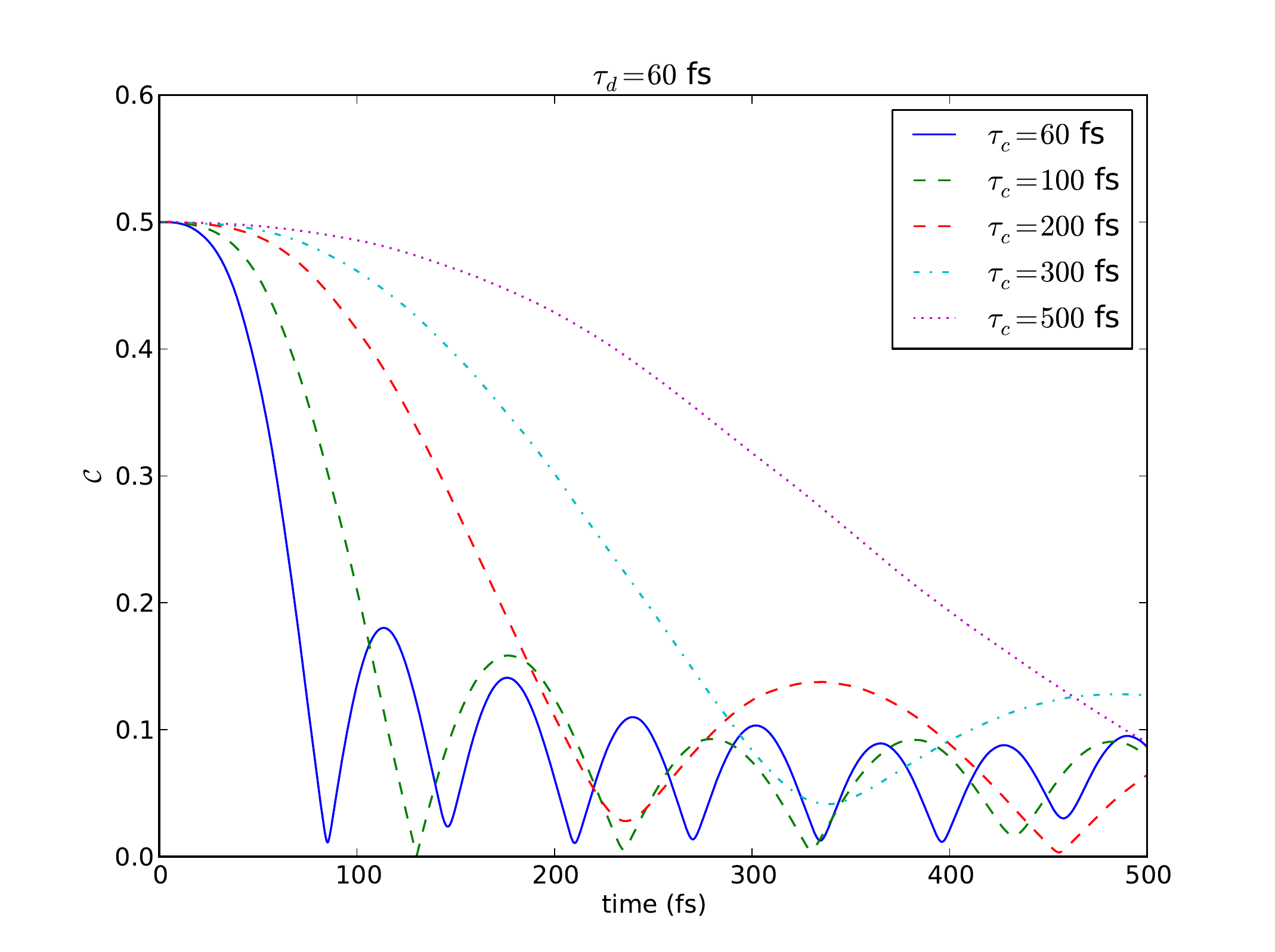}
\includegraphics[scale=0.39]{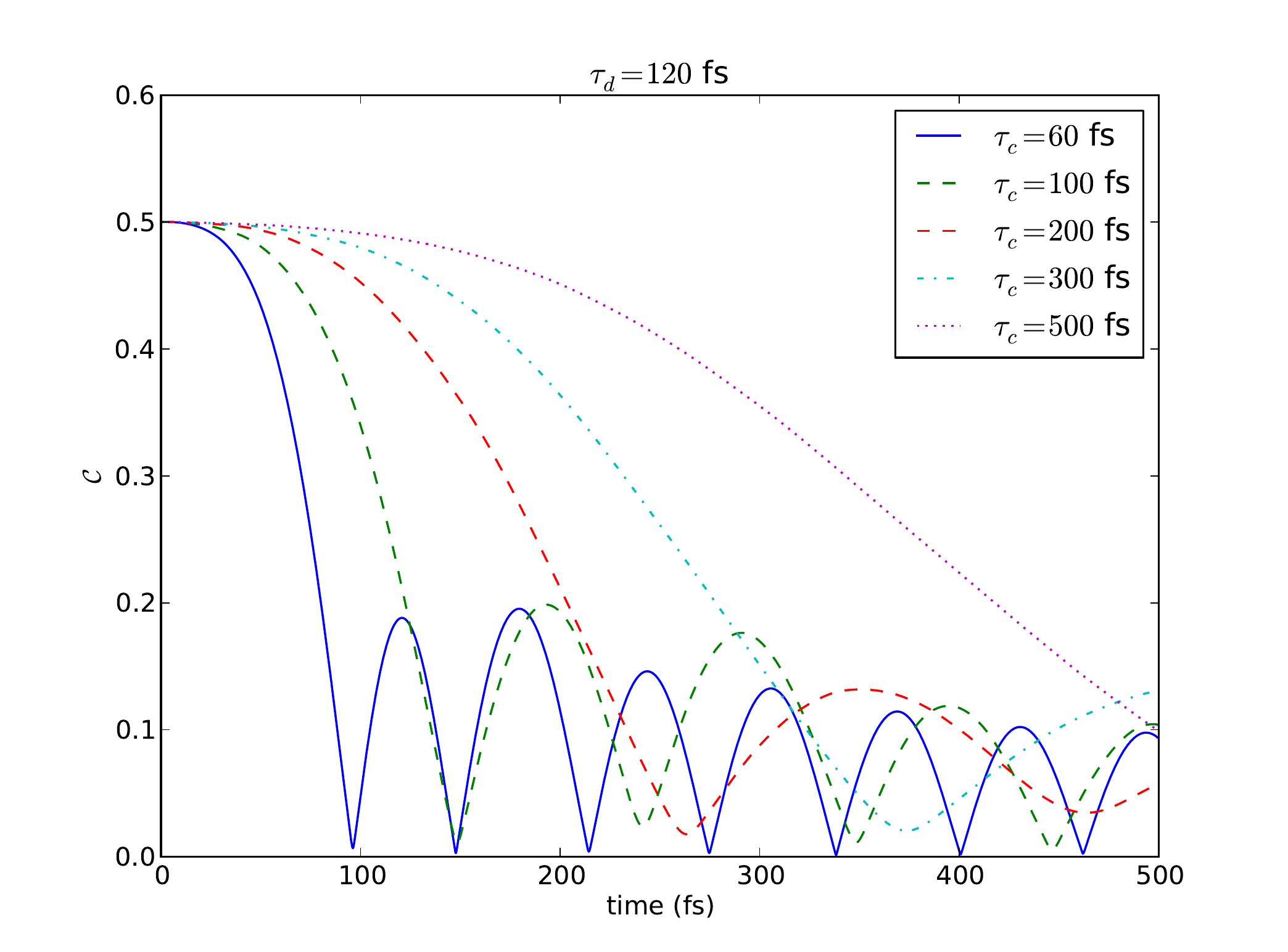}
\includegraphics[scale=0.39]{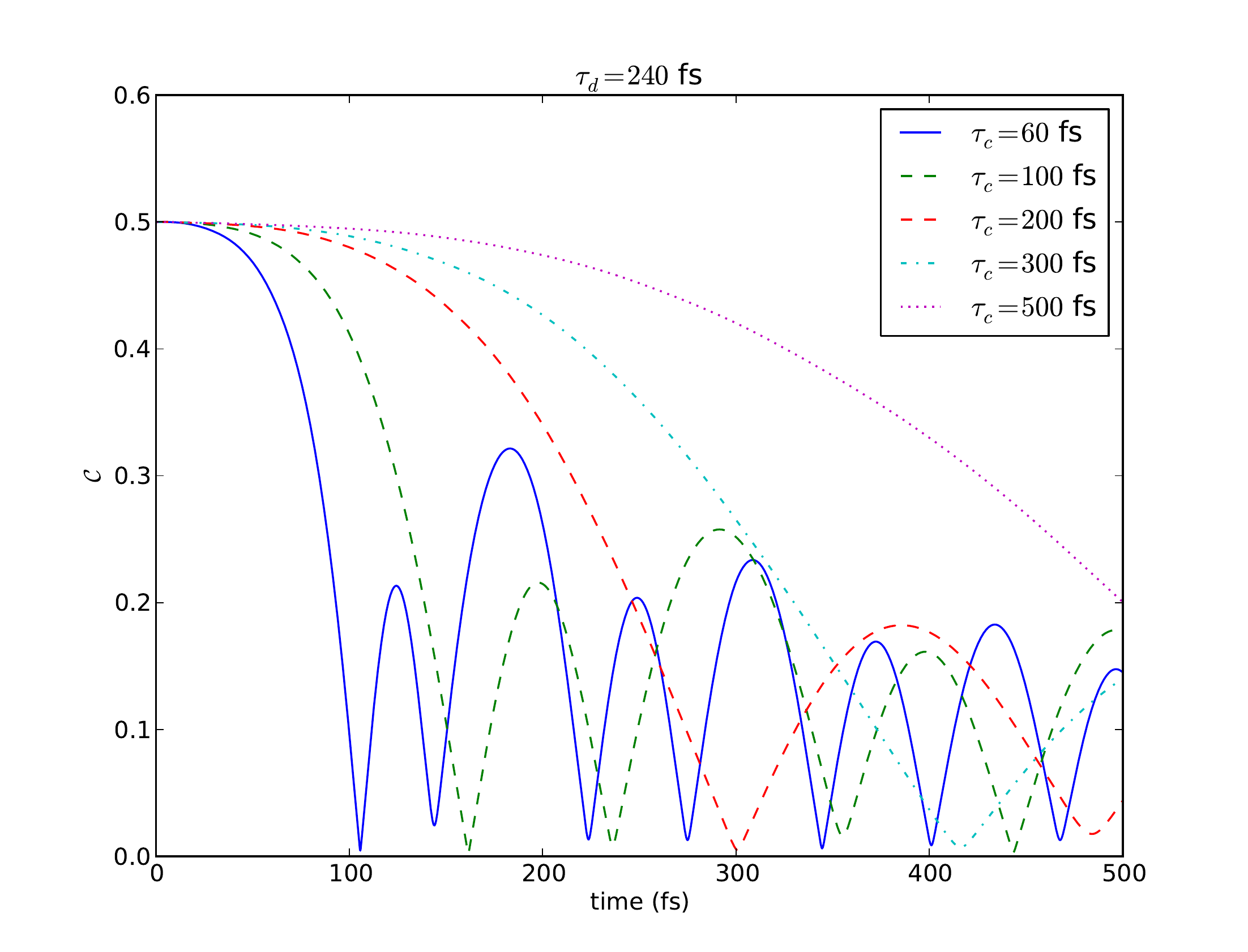}
\caption{$\mathcal{C}$ plotted against time for three  level
ladder system excited by thermally broadened CW source for various
level splittings at fixed $\tau_{d}$. The values of $\tau_{d}$
used are $\tau_{d} = 60$ fs (top), $\tau_{d} = 120$ fs (middle)
and $\tau_{d} = 240$ fs (bottom). For large values of time, the
quantity $\mathcal{C}$ becomes smaller and as  $t \rightarrow
\infty$ the value $\mathcal{C}$  approaches zero. }
\label{purmathcal}
\end{center}
\end{figure}

\begin{figure}[h]
\begin{center}
\includegraphics[scale=0.45]{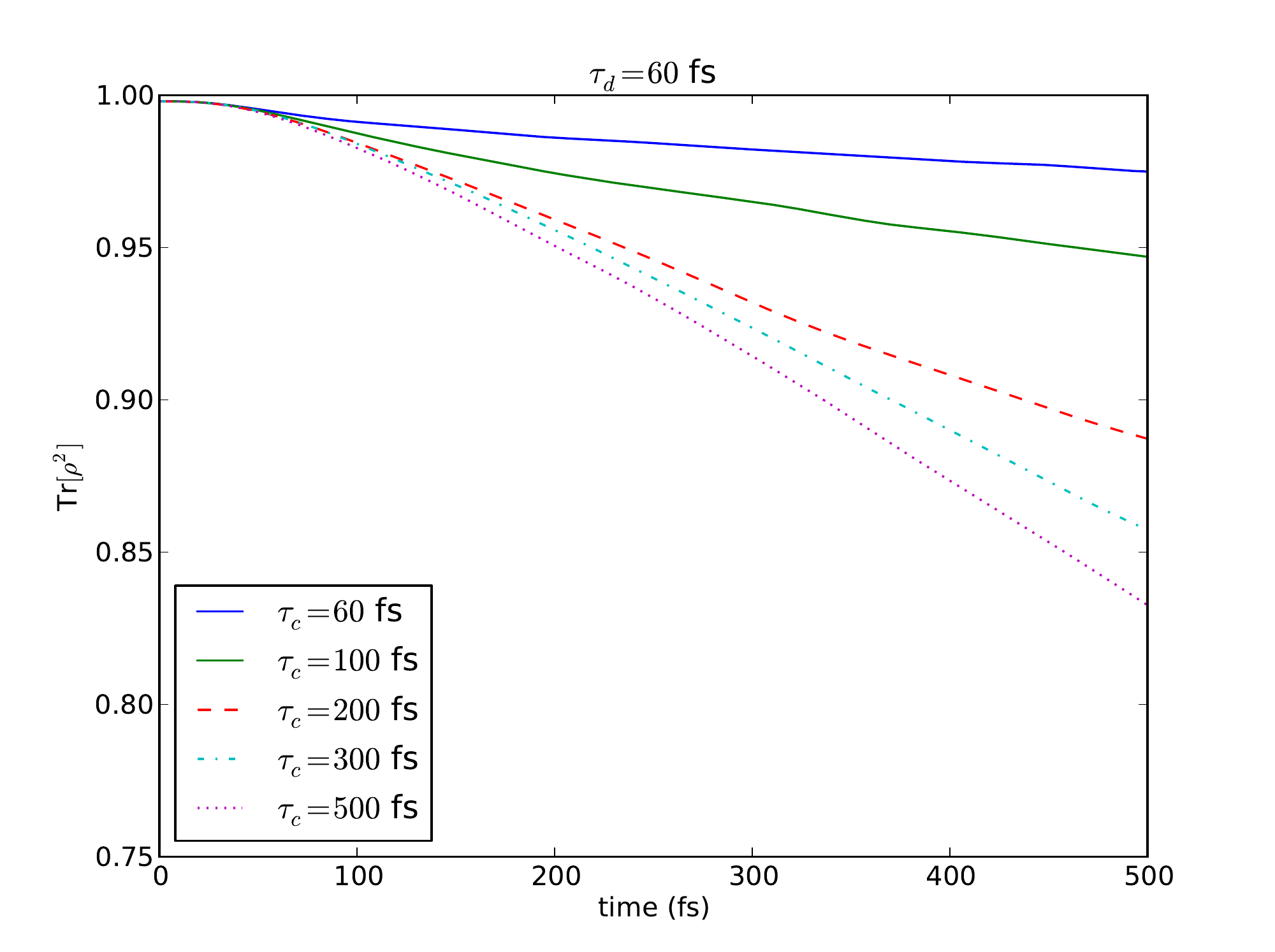}
\includegraphics[scale=0.45]{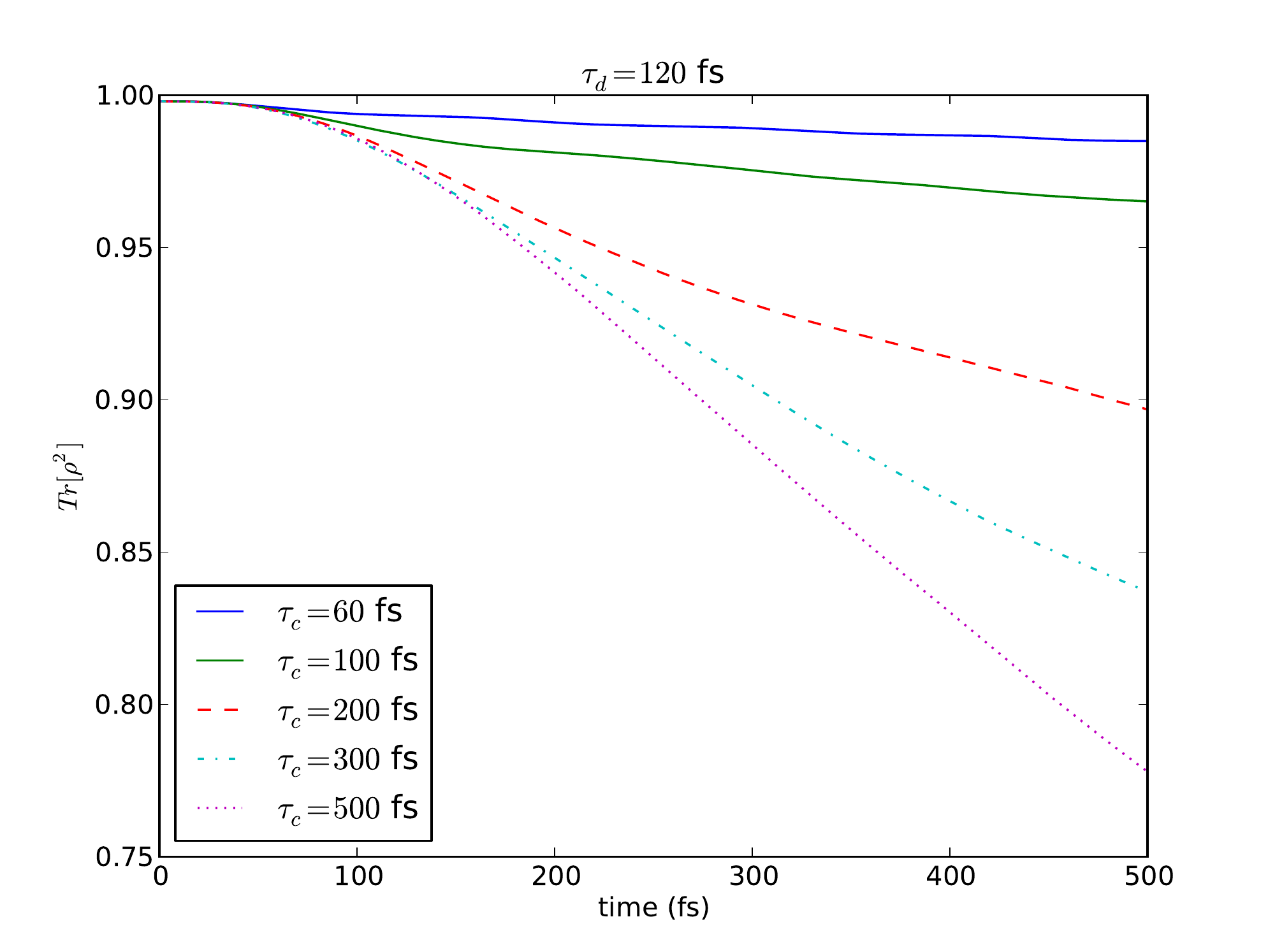}
\includegraphics[scale=0.45]{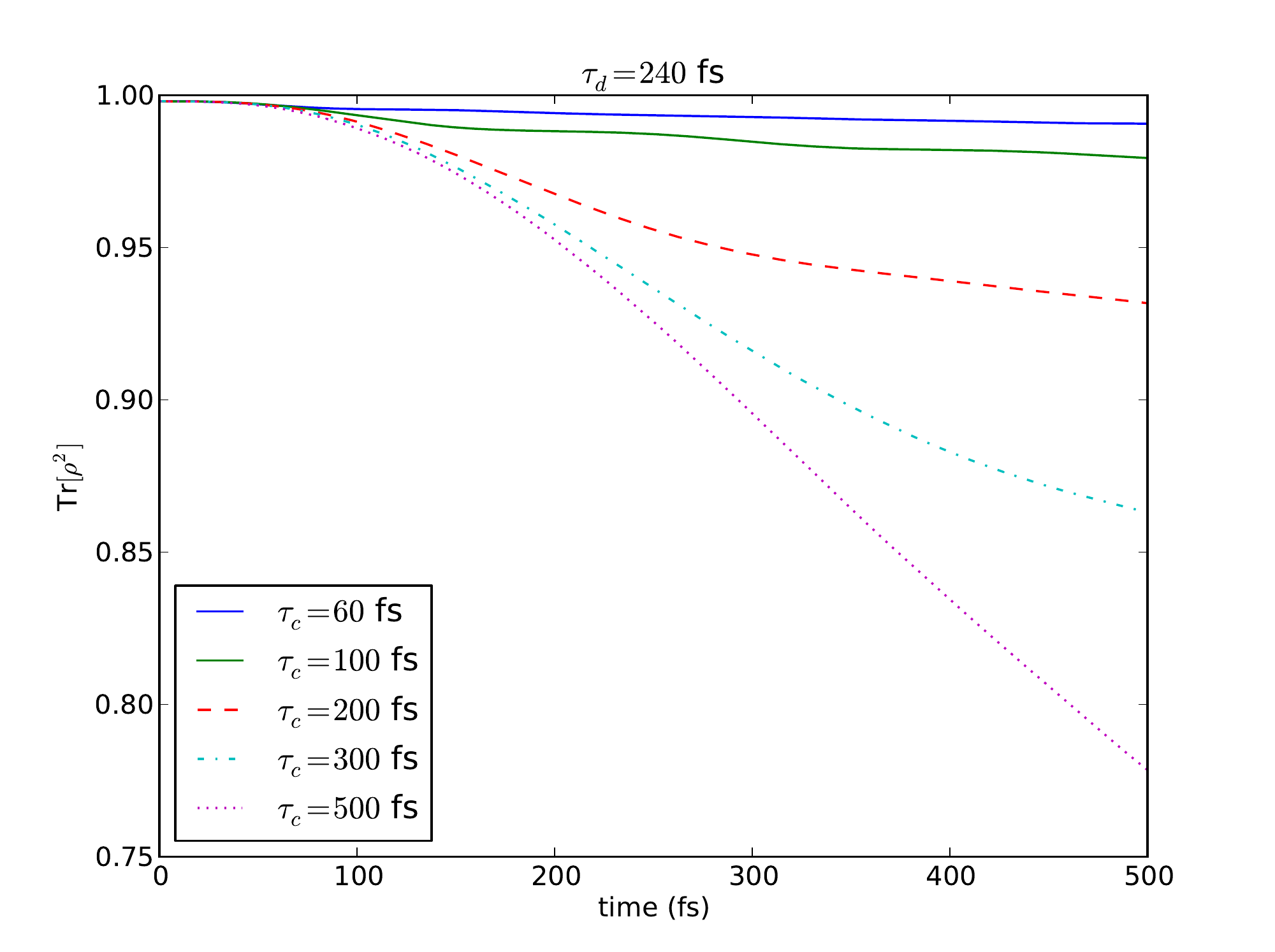}
\caption{System purity $\text{Tr}\left(\rho^{2}\right)$
plotted as a function of time for various excited state periods
$\tau_{c}$ and radiation coherence times $\tau_{d}$ for a three level ladder system excited by thermally broadened CW source. As the
excited state period $\tau_{c}$ becomes larger, the purity of the
system decreases at a faster rate. The following $\tau_{d}$ times
are shown: 60 fs (top), 120 fs (middle) and 240 fs (bottom). }
\label{pur}
\end{center}
\end{figure}

\begin{figure}[h]
\begin{center}
 \includegraphics[scale=0.4]{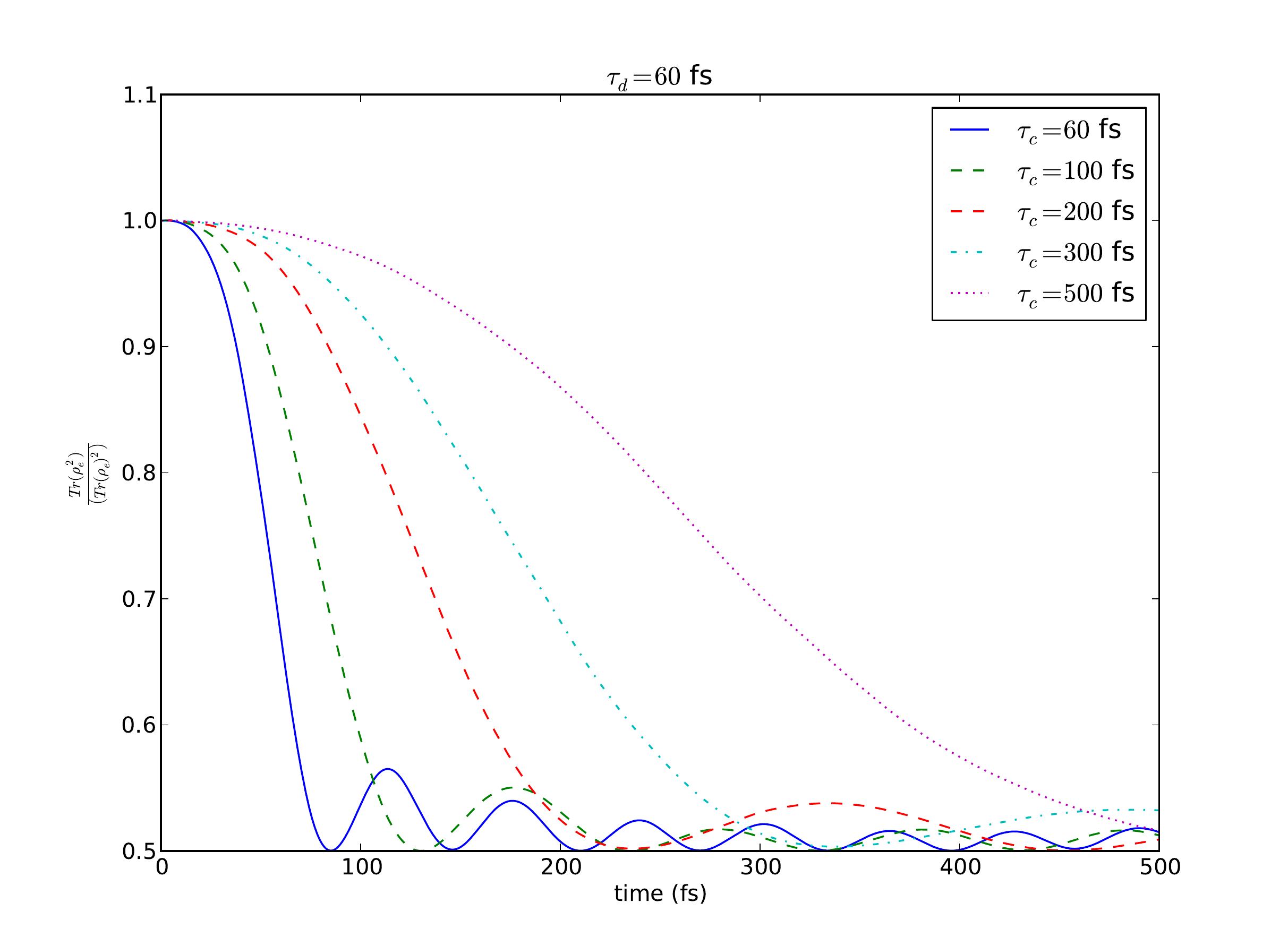}
\includegraphics[scale=0.45]{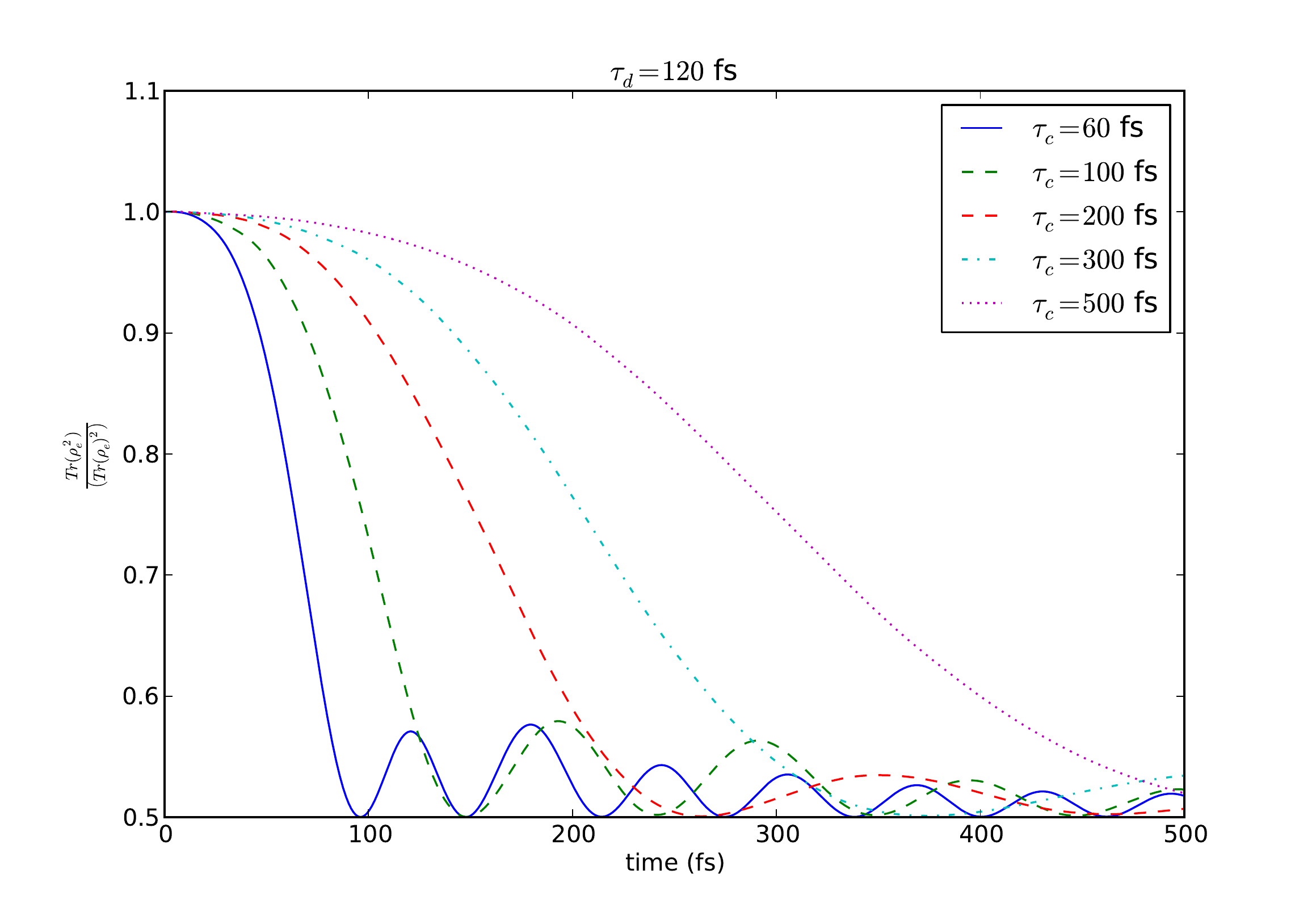}
\includegraphics[scale=0.5]{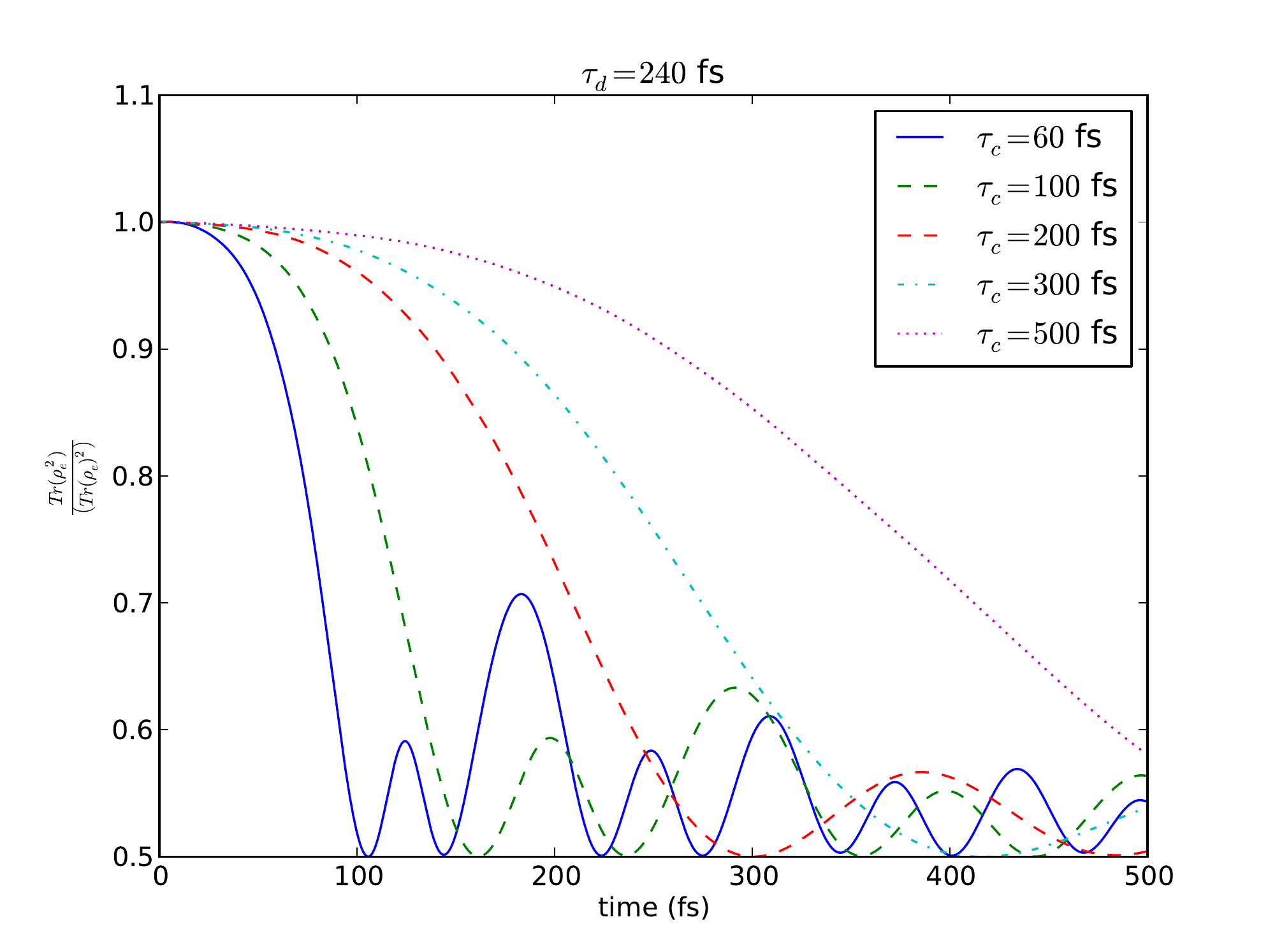}
\caption{Excited state purity plotted against time for
three level ladder system excited by thermally broadened CW source
for various level splittings, $\tau_{c}$ at various $\tau_{d}$. The
following $\tau_{d}$ times are shown: 60 fs (top), 120 fs (middle)
and 240 fs (bottom).  }
\label{pure}
\end{center}
\end{figure}

To examine the coherence of the created excited state, we define
\begin{equation} \mathcal{C}\equiv|\langle\rho_{23}\rangle|/(\langle\rho_{33}\rangle+\langle
\rho_{22} \rangle) \label{Cm}\end{equation}
 as a measure of the mixed state character;
here,  $\mathcal{C} = 0.5$ corresponds to that of an equal coherent
superposition of two states.  We plot this quantity, as well as the purity for
the entire system $\text{Tr}[\rho^2]$, in Figs. \ref{purmathcal}
and $\ref{pur}$ for several values of $\tau_{c}$ and $\tau_{d}$. To
complement our measure $\mathcal{C}$ we also plot the full system purity $\textrm{Tr}(\rho^2)$ in Fig. \ref{pur}  and the excited
state purity $\textrm{Tr}(\rho^{2}_{e})/(\text{Tr}^{2}(\rho_{e}))$
in Fig. \ref{pure}.

As seen in  Figs.   \ref{popcohcomp} and as manifest in Fig. \ref{purmathcal}, as the
system evolves under incoherent excitation,  coherences stay
the same order of magnitude while the populations increase.  In
the long time limit (not shown), we recover the  limit in Ref. \cite{brumer}
insofar as the state populations become overwhelmingly larger than
the coherences between them, i.e., $\mathcal{C}
\rightarrow 0$ in the limit of $t \rightarrow \infty$.  This time becomes longer with increasing
$\tau_{d}$  but is, of course, infinitesimally small in cases of
natural light, which illuminates  for macroscopic time scales.  Note that both $\mathcal{C}$ and the excited state purity  are found to
behave similarly and that the coherence is seen to survive longer for
larger $\tau_{d}$ and larger $\tau_{c}$. The
curves show decaying  oscillations at a frequency $2 \pi / \tau_{c}$.

\begin{figure}[h]
\begin{center}
\includegraphics[scale=0.4]{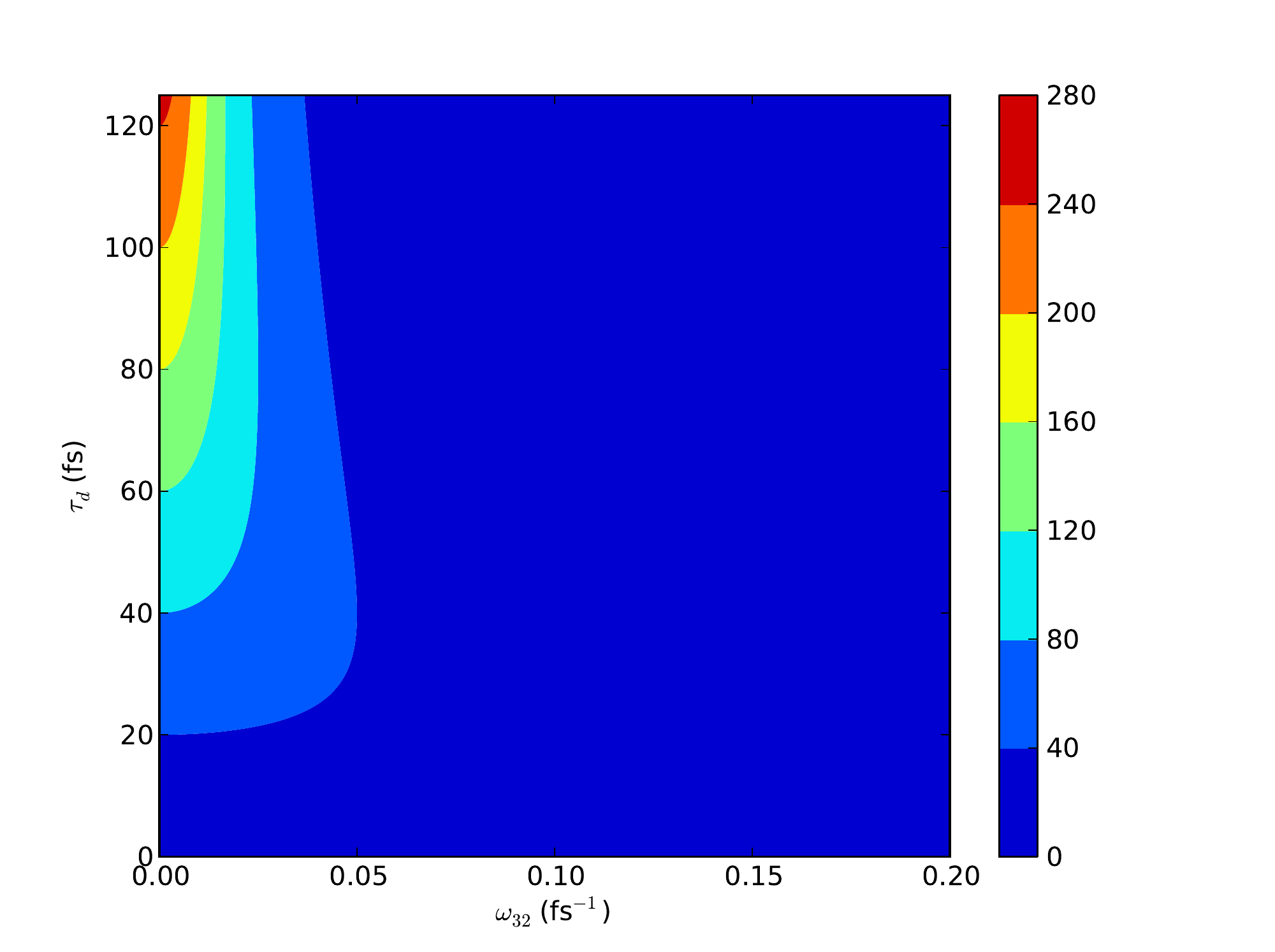}
\caption{Plot of $\mathcal{U}(\omega_{32},\tau_{d})$,  red regions
represent highest intensity and blue regions represent lowest
intensity.} \label{mathcalUp}
\end{center}
\end{figure}

Examining  the plot of $\text{Tr}[\rho^{2}]$ in Fig. \ref{pur} it is
evident that the total purity decreases with time, where  the smaller
$\tau_c$ is relative to $\tau_{d}$, the less pure the state becomes.
This effect is, however, mainly due to increased population in the excited state levels,
determined by $\mathcal{U}(\omega)$ [see Eq. (\ref{mathcalU})].
For example, when $\omega_{0}$ is in the center of the levels,
$\mathcal{U}= 8\tau_{d}/(4+\tau_{d}^2\omega_{32}^{2})$. By
plotting this as function of $\tau_{d}$ and $\omega_{32}$ (Fig.
\ref{mathcalUp}) it is clear that as the level spacing becomes
larger, the rate of pumping decreases, and vice versa.

\subsection{Toy  PC645 in a Collisionally Broadened CW Source}

\begin{figure}[h]
\begin{center}
\includegraphics[scale=0.5]{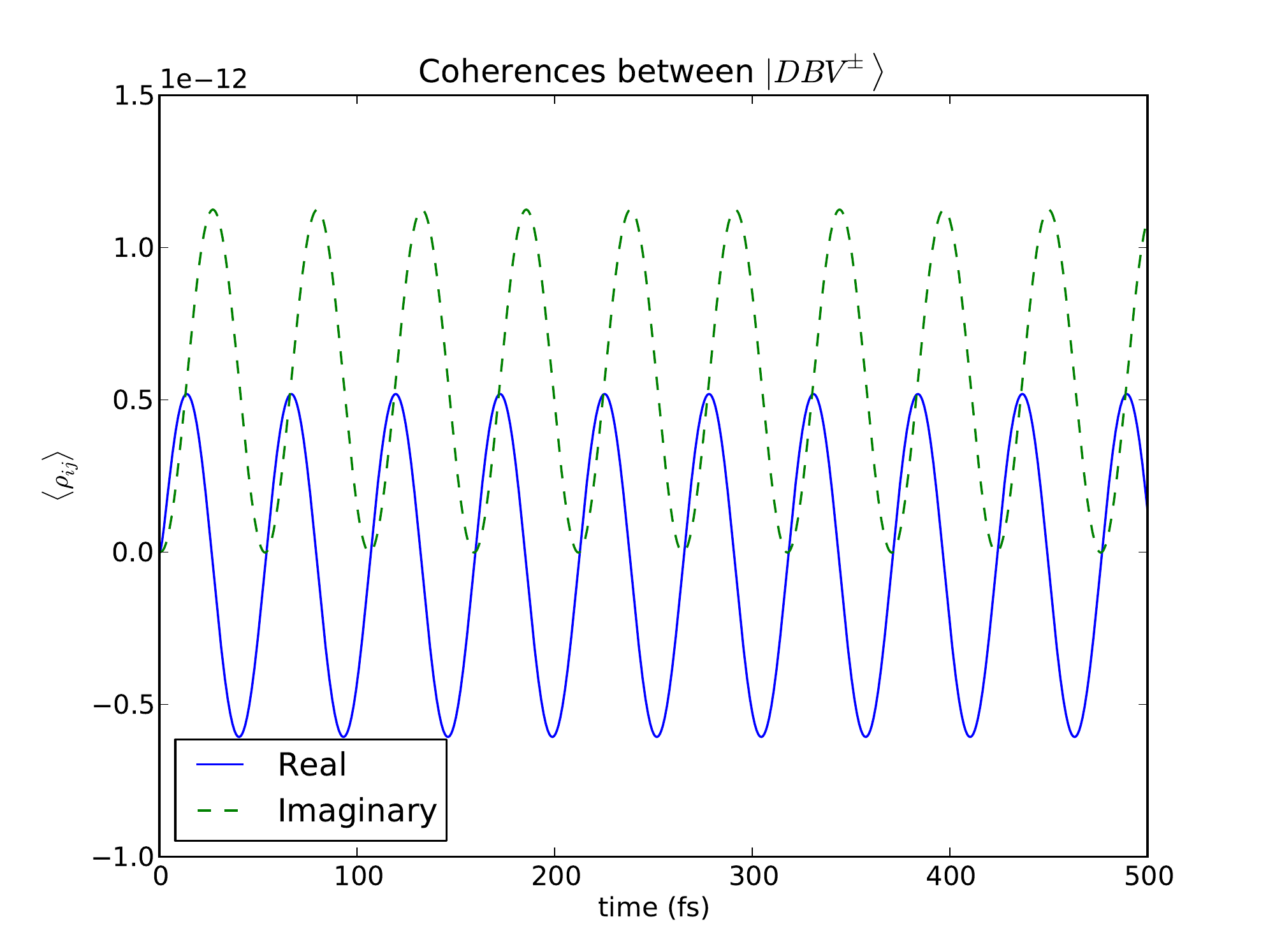}
\includegraphics[scale=0.5]{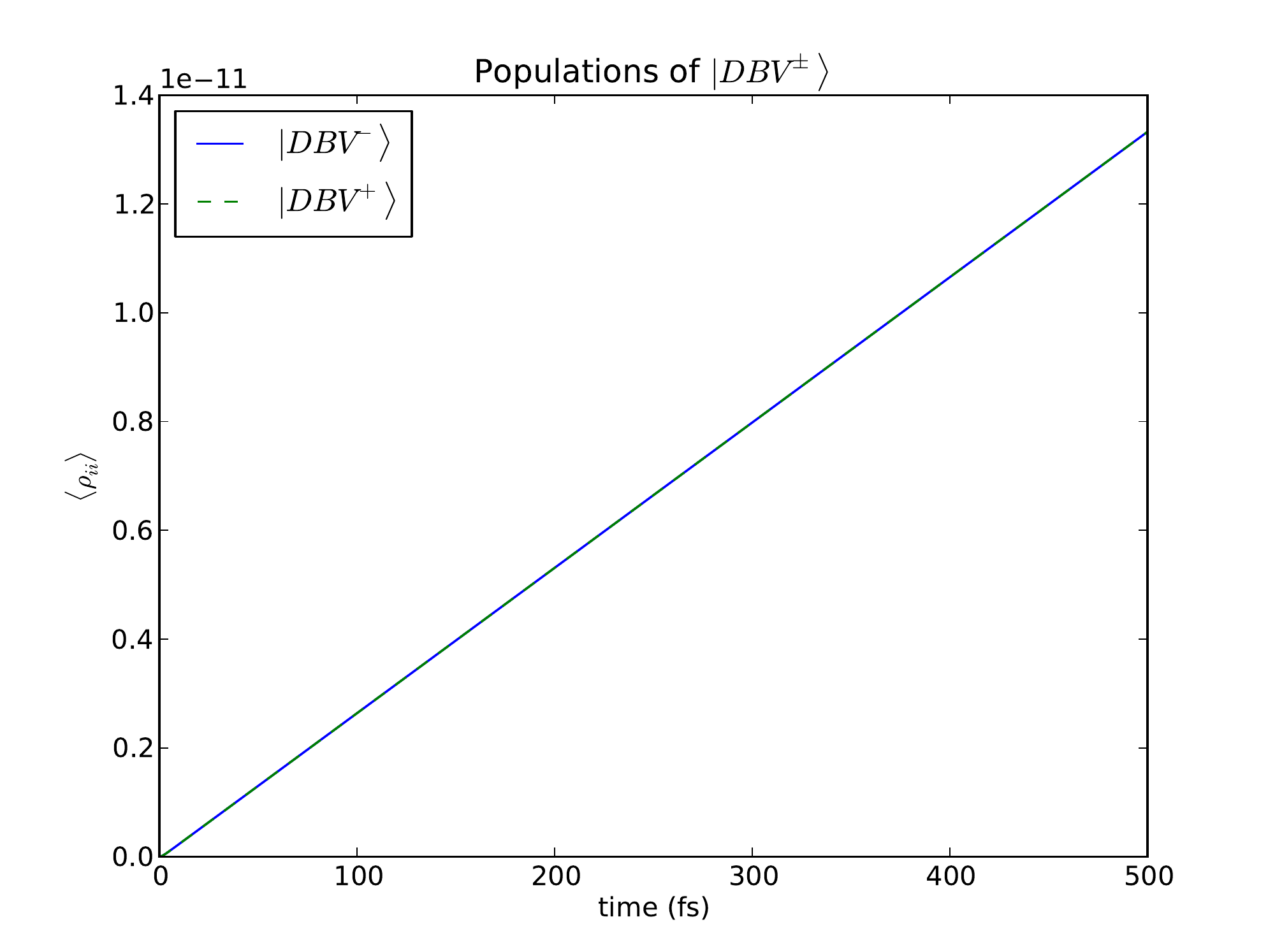}
\caption{ (Top) Coherences between the $|DBV^{-}\rangle$ and
$|DBV^{+}\rangle$ in the toy PC645 model as a function of time for the sudden turn-on case. (Bottom) Populations of
$|DBV^{-}\rangle$ and $|DBV^{+}\rangle$ in the toy model of PC645  as a function of time for the sudden turn-on case.
Excitation frequency of the laser is in the middle of the two
transitions.} \label{pc645cwcoh}
\end{center}
\end{figure}

\begin{figure}[h]
\begin{center}
\includegraphics[scale=0.5]{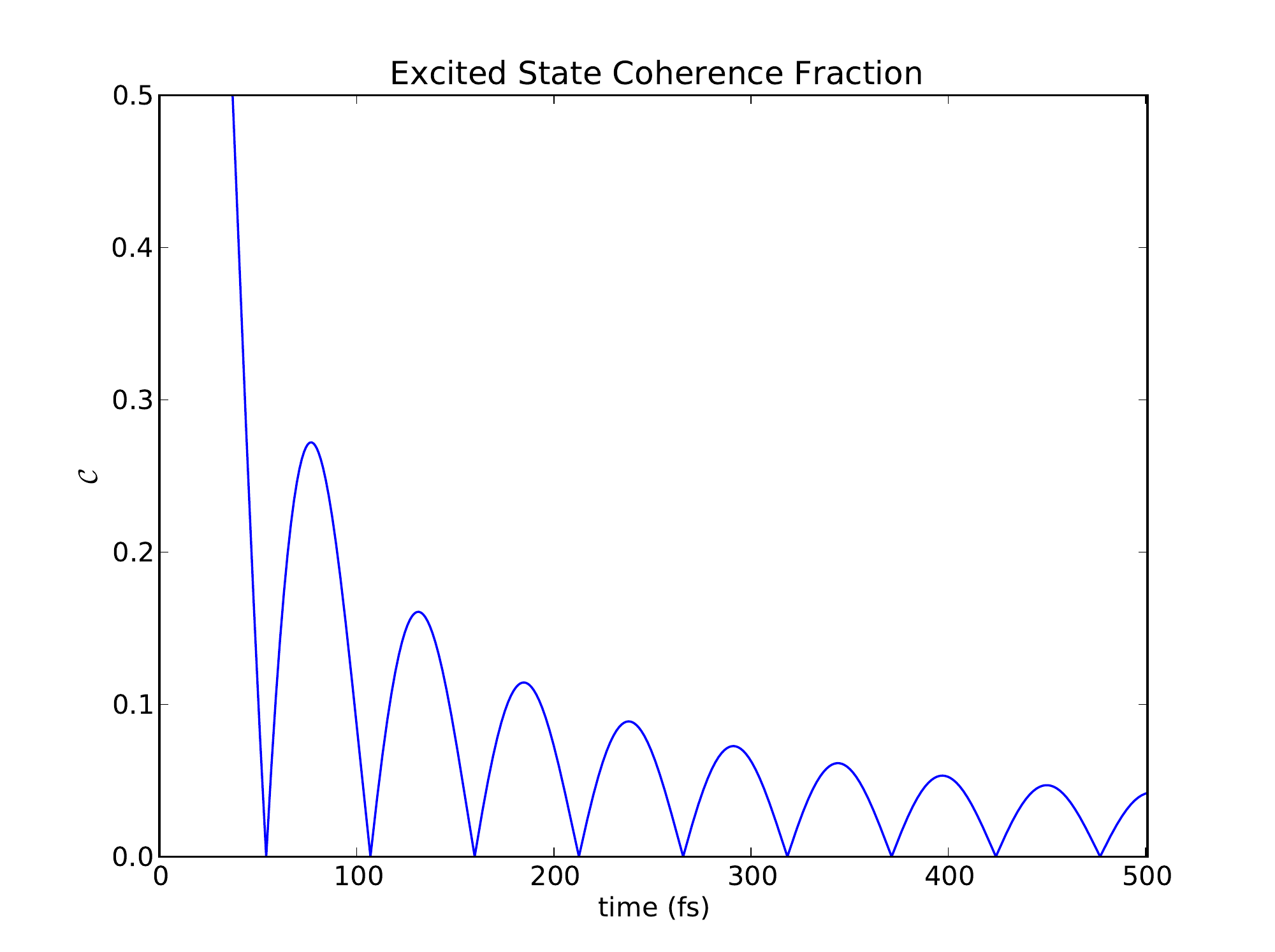}
\caption{Plot of $\mathcal{C} \equiv
\frac{|\rho_{23}|}{\rho_{22}+\rho_{33}} $ for a toy model of  PC645 upon
excitation by a collisionally broadened CW source, for parameters
indicated in the text.} \label{pc645cwC}
\end{center}
\end{figure}

Intense interest has surrounded the observation of quantum
coherence in photosynthetic pigment protein complexes \cite{scho},
observed with coherent laser light. However, under natural
conditions, these protein complexes are irradiated with sunlight.
Here we utilize  $V$ level system as a toy model for the
electronic energy levels of PC645 and focus is on obtaining
qualitative insight into the time scales for  coherences
due to coupling to the chaotic light. The upper levels $|2\rangle$
and $|3\rangle$ represent $|DBV^{-}\rangle$ and $|DBV^{+}\rangle$
electronic states that are excited experimentally
\cite{scho,sch1}. This model PC645 level structure is irradiated
with the collisionally broadened CW source using the perturbative
approach described above.  The laser frequency $\omega_{0}$ was
chosen to be in the center of the two levels so as to excite both
transitions equally and the  coherence time $\tau_{d} = 1.32$ fs,
is that of sunlight. Electric field intensities are taken from
literature values of the intensity of the sun at midday
\cite{sunvalues}. Site energies and dipole moments given in
Appendix D were taken from literature \cite{sch1}. The
perturbative calculation for the coherences and populations of the
excited states, where the field is turned on at $t_0 = 0$, are
presented in Figs. \ref{pc645cwcoh} and \ref{pc645cwC}.

Once again from Figs. \ref{pc645cwcoh} and \ref{pc645cwC} it is
evident  that there is a transient coherent response associated
with the sudden turn-on that becomes an increasingly smaller
fraction of the excited state population, as evidenced by the
measure $\mathcal{C}$. For example, for $t > 500$ fs, $\mathcal{C}
< 0.05$, i.e., the coherences are already very small relative to
population.  Since the population growth is linear in time,
$\mathcal{C}$ decays to zero  as $|\sin{(\omega_{23}t)}|/(2t)$.
 Note, similar to the computations above, the timescale over which
the coherences are a significant fraction of the populations is
far larger than $\tau_d$.  This is consistent with our observation
that this timescale increases as $1/\tau_c$.

This calculation was done on an isolated quantum system,  without
the inclusion of external degrees of freedom corresponding to  the
local vibrations and the protein environment. Depending on the
parameters of the bath, the decoherence associated with
system-bath interactions could speed up the decoherence
experienced by interacting with light or have negligible impact.
Since excitation from pigment protein complexes need tens of
picoseconds to arrive at the reaction center,  it  is clear, even
from this toy model,  that it cannot do so via pure electronic
coherent dynamics.

These computations assume the sudden turn-on of the incoherent
light, which induces strong coherences.  The issue of a slow
turn-on is addressed in the following section.

\section{Response of an Isolated Quantum System to Long Incoherent Pulses}

 A potential issue associated with  using the collisionally broadened  CW source discussed
  above is the ``sudden turn-on effect" which will induce artificial
coherences. This issue  can be examined  by using a ``noisy pulsed
source"   \cite{brumerbook},  with a two time correlation function
given in Eq. (\ref{corr2}). However, using the perturbative
approach outlined in Sec. IIA only gives analytic results only for
times $t\gg\tau_{p}$, where $\tau_{p}$ is the pulse duration.
Specifically, inserting Eq. (\ref{corr2}) into Eq. (\ref{den1})
and setting $t_{0} \rightarrow -\infty$ and $t \rightarrow \infty$
gives
\begin{equation}
\rho_{e}\left(t\gg\tau_{p}\right)=\sum_{i,j}c_{i}c_{j}^{*}|E_{i}\rangle\langle E_{j}|e^{-i\omega_{ij}t}\eta_{P}^{ij}
\label{npden}
\end{equation}

\begin{equation}
\eta^{ij}_p = \tau_pTe^{it_m \omega_{ij}} \exp \left(- \frac{\tau^2_p \omega^2_{ij}}{8} \right) \exp\left(-\frac{T^2}{8}(\delta_i + \delta_j)^2 \right) \end{equation}
where $T = \tau_p\tau_d/\sqrt{\tau^2_p + \tau^2_d}$

\begin{figure}[h]
\begin{center}
\includegraphics[scale=0.5]{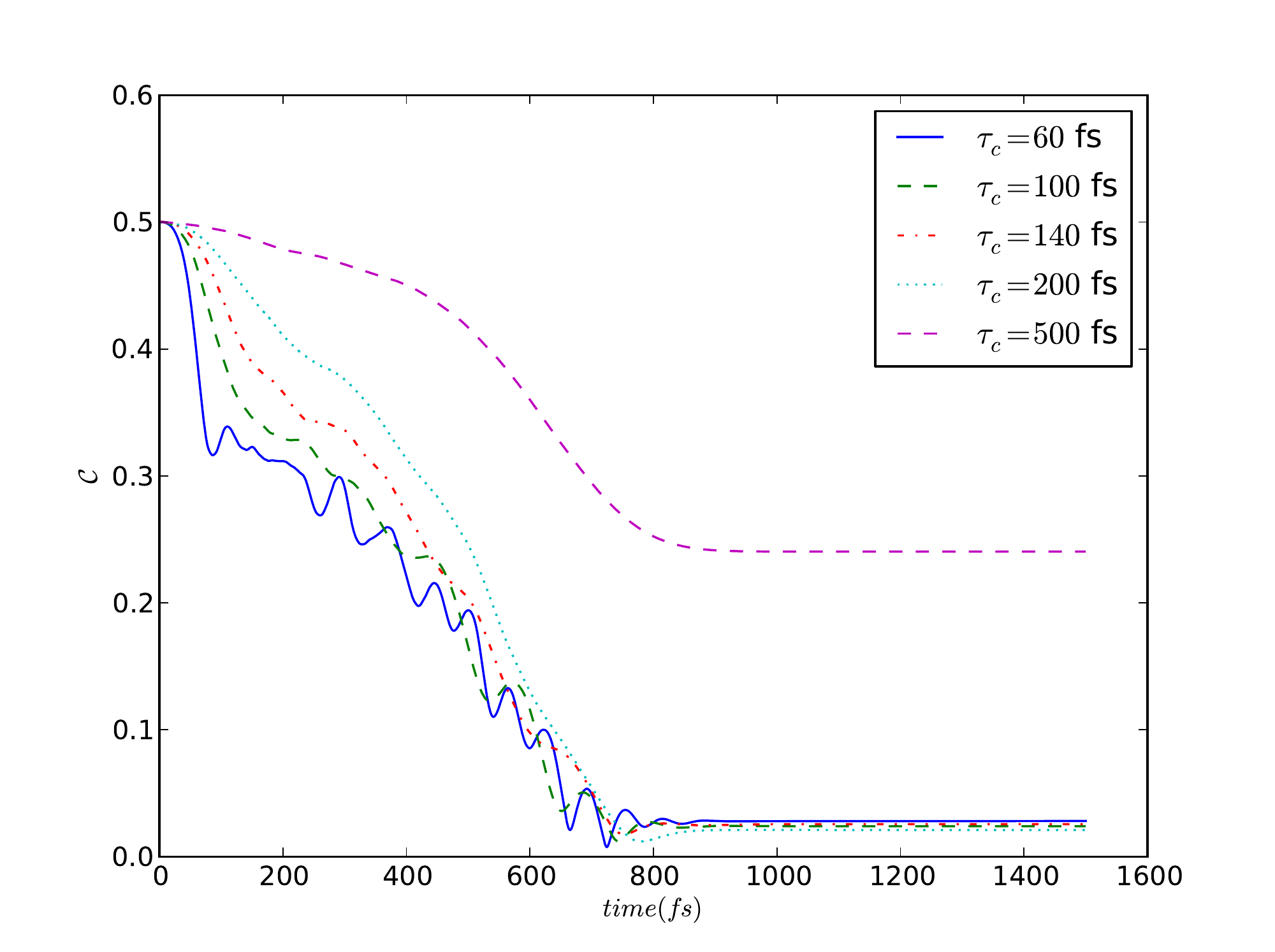}
\caption{Plot of $\mathcal{C}$ for various
excited state splittings, $\tau_{c}$ for long incoherent pulses incident on a three level ladder system.  The pulse used is $1$ ps in duration. $\tau_{d} = 120$ fs. The frequency center of
the pulse is chosen to be in the center of the two transitions.}
\label{npmathcalC}
\end{center}
\end{figure}

Note that in the limit of $\tau_p \gg \tau_d$, $T \rightarrow \tau_d$.   Further note that, unlike the sudden turn-on case, averaging over pulse centers $t_m$ would cause coherences to vanish.  Using Eq. (\ref{npden}) gives a  general form for $\mathcal{C}$ between states $i$ and $j$ of
\begin{equation}\mathcal{C} = \frac{\exp(-\tau^2_p \omega^2_{ij}/8) \exp(-T^2(\delta_i + \delta_j)^2/8)}{\exp(-T^2\delta^2_i/2) + \exp(-T^2\delta^2_j/2)}\end{equation}
For the specific case where the $\omega_0$ lies between the two eigenvalues, $\delta_i = -\delta_j = \omega_{ij}/2$.  Hence, in this case, \begin{equation}
\mathcal{C} = \frac{1}{2} \exp(-\tau^2_p \omega^2_{ij}/8) \exp(T^2 \delta^2_i/2) \label{eq16} \end{equation}
which, if $\tau_p \gg \tau_d$, becomes
\begin{equation}
\mathcal{C}=  \frac{1}{2} \exp\left[ (\tau^2_d - \tau^2_p)~\omega^2_{ij}/8\right] = \frac{1}{2}\textrm{ e}^{-\tau^2_p \omega^2_{ij}/8} \label{eq18}\end{equation}
The latter expressions assume that the
dipole moment of the excited states are equal. We have performed
numerical calculations on such systems and these results, which
allow for an appreciation of the coherences generated during the
pulse,  are presented in Fig. \ref{npmathcalC} where $\tau_p = 1$ ps. During the pulse,
the excited state coherences can be seen to be a significant
fraction of the excited state population.  However, after the
pulse is over, these coherences become a negligible fraction of the
populations.  Note that as $\tau_{c}$ becomes larger, the coherent response during the pulse
and post-pulse becomes larger. For $\tau_{c}$ approaching
$\tau_{p}$ the decoherence is significantly less than when
$\tau_{c}$ is considerably smaller than $\tau_{p}$.  Indeed, for $\tau_p = 1$ ps and $\tau_c = 500$ fs $\mathcal{C}$ is seen to be constant at 0.24 after the pulse is over, since both populations and coherences are constant after that time.  

These results are in contrast to those  of the collisionally
broadened CW source insofar as the time $\tau_d$ plays little role.  Rather, from the perturbative
expression [Eq. (\ref{eq18})] $\mathcal{C}$ is seen to be dependent
only on the ratio of pulse duration and excited state splitting.
Hence, even for incoherent sources with very small
$\tau_{d}$  the pulse  can create a partially coherent
superposition between excited states.  This is particularly the case for levels that are very closely spaced, e.g., vibrational levels in small molecules.   This is consistent with
observations made in Ref. \cite{brumer}, where note was made of
the fact that such coherence is generated because the
\textit{envelope} of the pulse is, itself, smooth.

\subsection{Irradiation of Toy PC645 Using Noisy Pulses}
\begin{figure}[h]
\begin{center}
\includegraphics[scale=0.5]{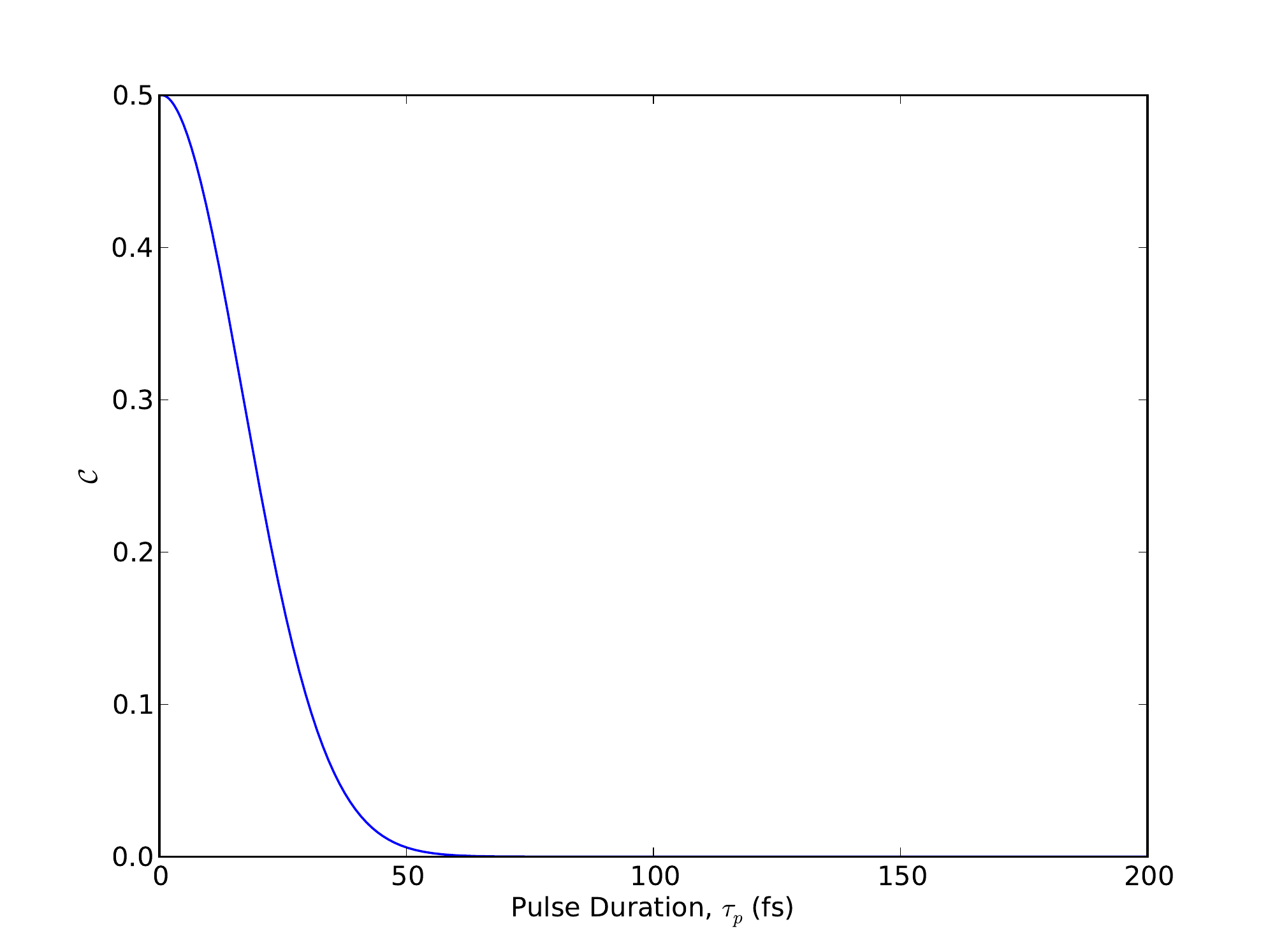}
\caption{$\mathcal{C}$, the magnitude of the post pulse excited
state coherence fraction of a toy model of PC645, plotted as a function of
pulse duration, $\tau_{p}$. For  long pulses, it is evident that
post pulse coherence is extremely small.} \label{pc645gam}
\end{center}
\end{figure}

As a demonstration of the response of electronic levels using typical biological system parameters, 
 consider the irradiation of a toy PC645 molecule with a noisy pulse. The model for PC645 is the same as in Sec. IIC.  The post pulse expression for $\mathcal{C}$, plotted as a function
of pulse duration, is shown in Fig. \ref{pc645gam} where $\tau_d <
\tau_p$ is not assumed.  For short pulses (sub 100 fs) partially coherent excited states
can be created, but for pulses longer than 100 fs the coherence is
negligible.   Hence, excitation of electronic superpositions using
incoherent sunlight of macroscopic time scale duration, is not
expected.

\section{Summary}

Understanding  the  response of molecular systems to incoherent
light,  such  as sunlight, is vital in efforts to
advance  studies of natural light harvesting processes,
photovoltaics, etc. Previous work 
\cite{brumer,onephoton,mancal} showed  that
the  density matrix of the molecule in the
long  time limit was that of a mixed state, with no evident time
dependent quantum coherence.

This paper has carefully examined the
nature  of  the transient coherences associated with molecular excitation
with  incoherent  light  for  two  paradigmatic cases,  the sudden turn-on of
collisionally broadened CW light, and the excitation by a Gaussian
pulse with phase jitter. The role of the decoherence time of the
radiation $\tau_d$, the
pulse  duration $\tau_p$,  and  the  system timescale as measured by the
inverse of the energy level spacing $\tau_c$, were examined.

For  the  case  of  sudden turn-on of collisionally broadened
light,  time  dependent  coherences,  although persistent, were
found  to  become insignificant relative to the populations as
time  involved. Both  larger $\tau_d$. and larger $\tau_c$ enhanced
the timescales of these coherences. Averaging over the start time of
the  sudden  turn-on  resulted  in  the survival of time independent (stationary)
coherences that were larger for larger $\tau_c$.

In  the  pulsed  case,  the  ratio  of  the  coherences  to the
population is, for times long after the pulse is over, and where
$\tau_p >> \tau_d$,  heavily determined by
the ratio of the pulse duration to the level spacing, i.e. $\tau_p/\tau_c$.
Specifically, the larger this ratio, the greater the decoherence.
Qualitatively, the $\tau_p$
dependence arises from the fact that the
longer the pulse is on, the larger the population in
the  excited  levels.  As  noted  earlier in \cite{brumer} molecules
irradiated  by pulses with smooth envelopes,  even  if  there  are  phase jumps,
pick up  coherence  from  the  smooth pulse envelope. Hence, results are
expected to differ for models that employ, e.g.,  erratic pulse amplitudes.

Both cases show that the mixed state comes about in a fashion distinctly
different than that in scenarios 
typically used to explore decoherence (see, e.g.,
Refs. \cite{Elran1, Elran2}). In those cases the effect of an environment
on an initial superposition state, with no external driving field, shows
a characteristic decay of coherences, i.e. the decay of off-diagonal 
$\rho_{ij}$ of the system density matrix. In the case studied here, these
off-diagonal elements do not decay, but the ratio of these elements to
the populations decrease as time goes on, since the populations increase due to
the external driving field. Interestingly, despite these differences, the
two scenarios do share the common feature, i.e. that the approach to a mixed
state is slower for smaller level spacings, i.e., large $\tau_c$.
Hence, extended studies on incoherent light excitation in dense vibronic
manifolds in large molecules is certainly well motivated, and is
underway\cite{grinev}.

The fact that, in both sudden turn-on and pulsed cases, larger $\tau_c$
enhances coherence times indicates the need to consider decoherence effects
of the environment, if the system is open. Such environmental decoherence effects
may well serve as the dominant decoherence effect (as opposed to the effect
of the incoherent light) for systems with large $\tau_c$,
a resultant consistent with some earlier considerations  \cite{leo} in a
different context.
Further studies of this type on realistic open atomic and molecular systems are in progress.

\acknowledgments ZS thanks OGSST for funding,
and Dr. L. Pachon, Y. Khan and R. Dinshaw for edifying discussions.  PB thanks Dr. T. Tscherbul for insightful remarks.  
This work was supported by  the Air Force Office of Scientific Research under Contract No.  FA9550-13-1-0005

\begin{figure}[h]
\begin{center}
\includegraphics[scale=0.5]{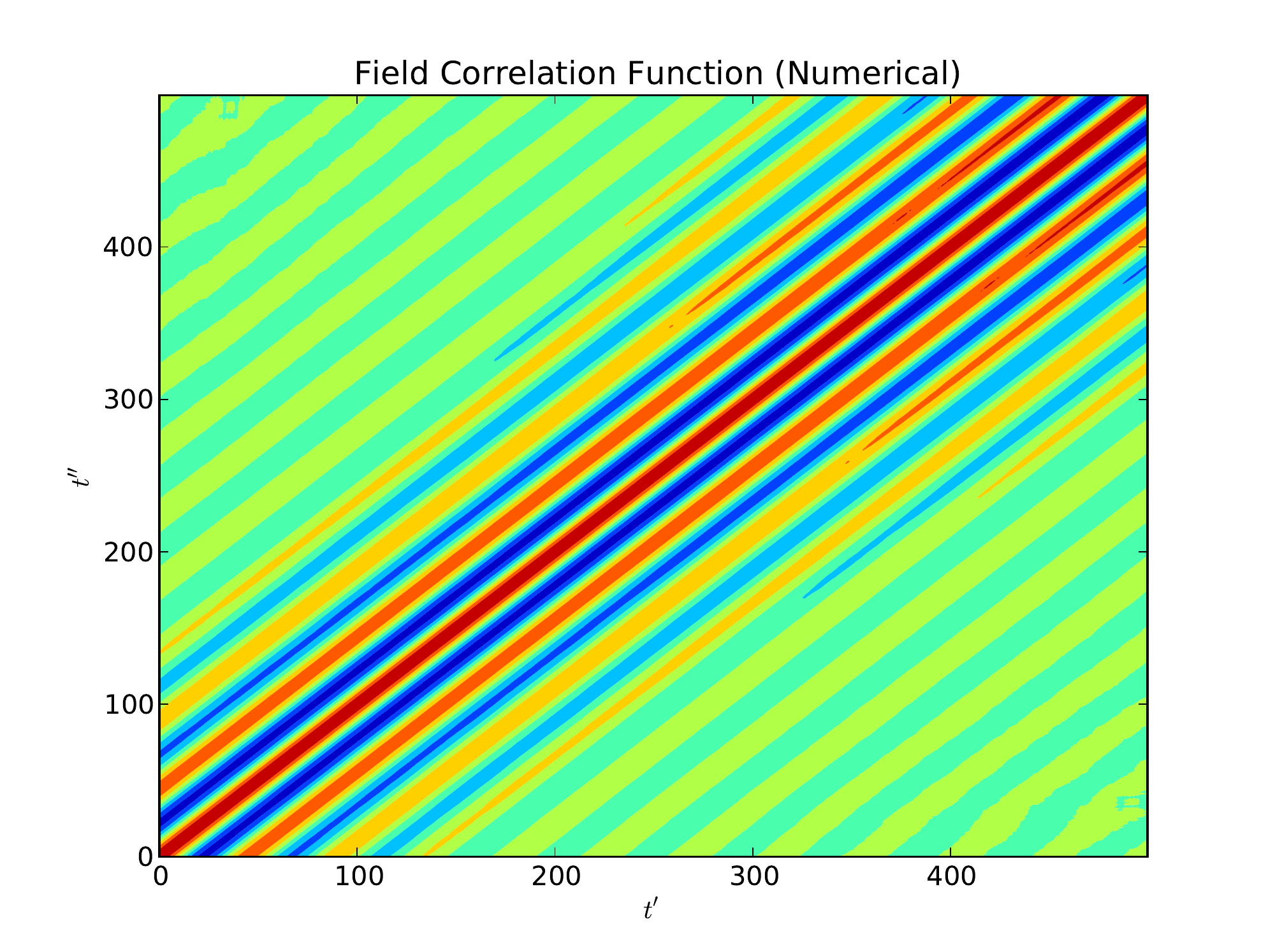}
\includegraphics[scale=0.5]{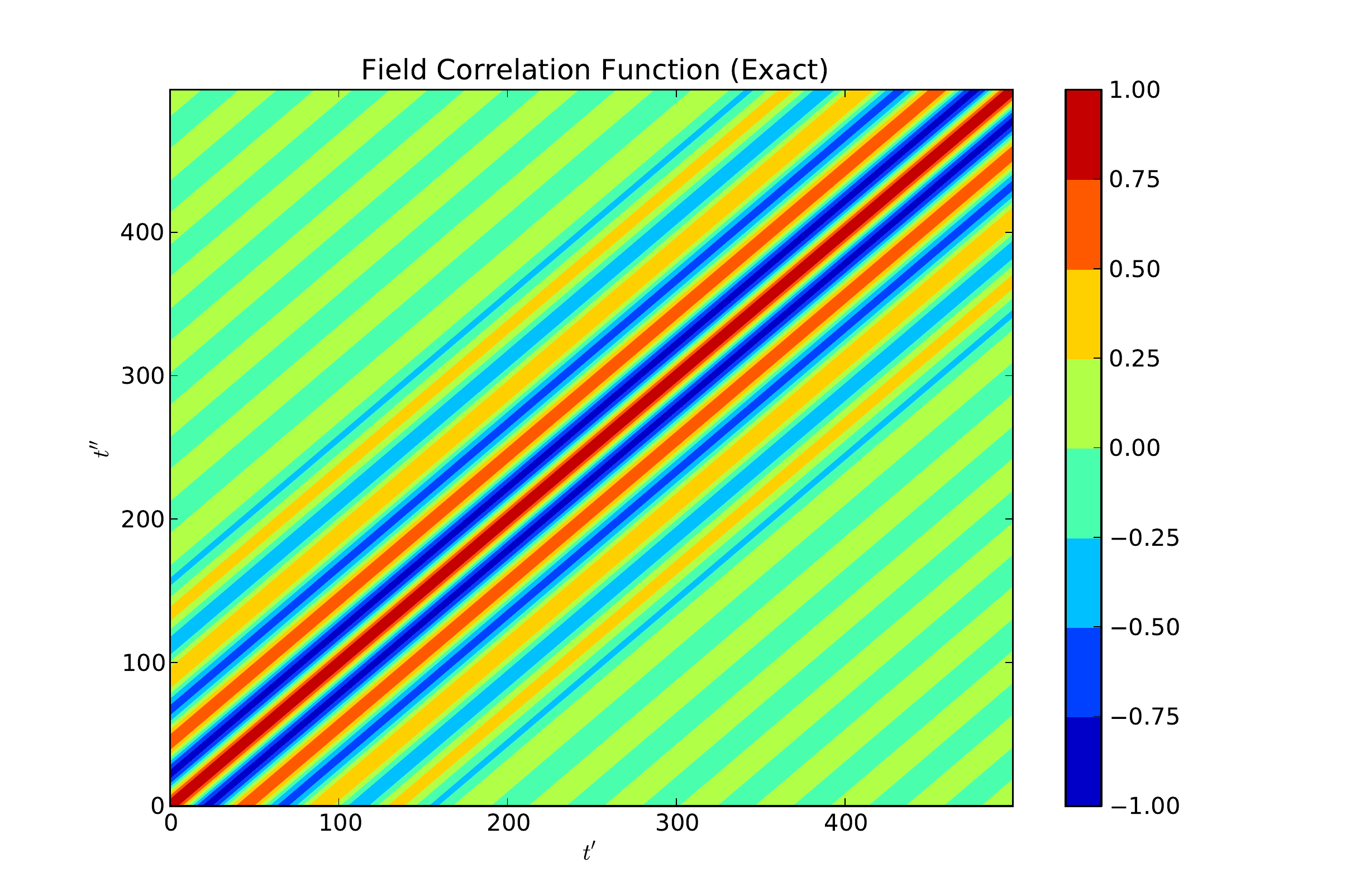}
\caption{  Comparison of numerically generated (top) and numerical
[Eq. (\ref{corr1})] (bottom)
$\langle\varepsilon(t^\prime)\varepsilon(t^{\prime\prime})\rangle$.
Both plots share the same color legend. The times are in units of
femtosecond.} \label{CWexnum}
\end{center}
\end{figure}

\begin{figure}[h]
\begin{center}
\includegraphics[scale=0.5]{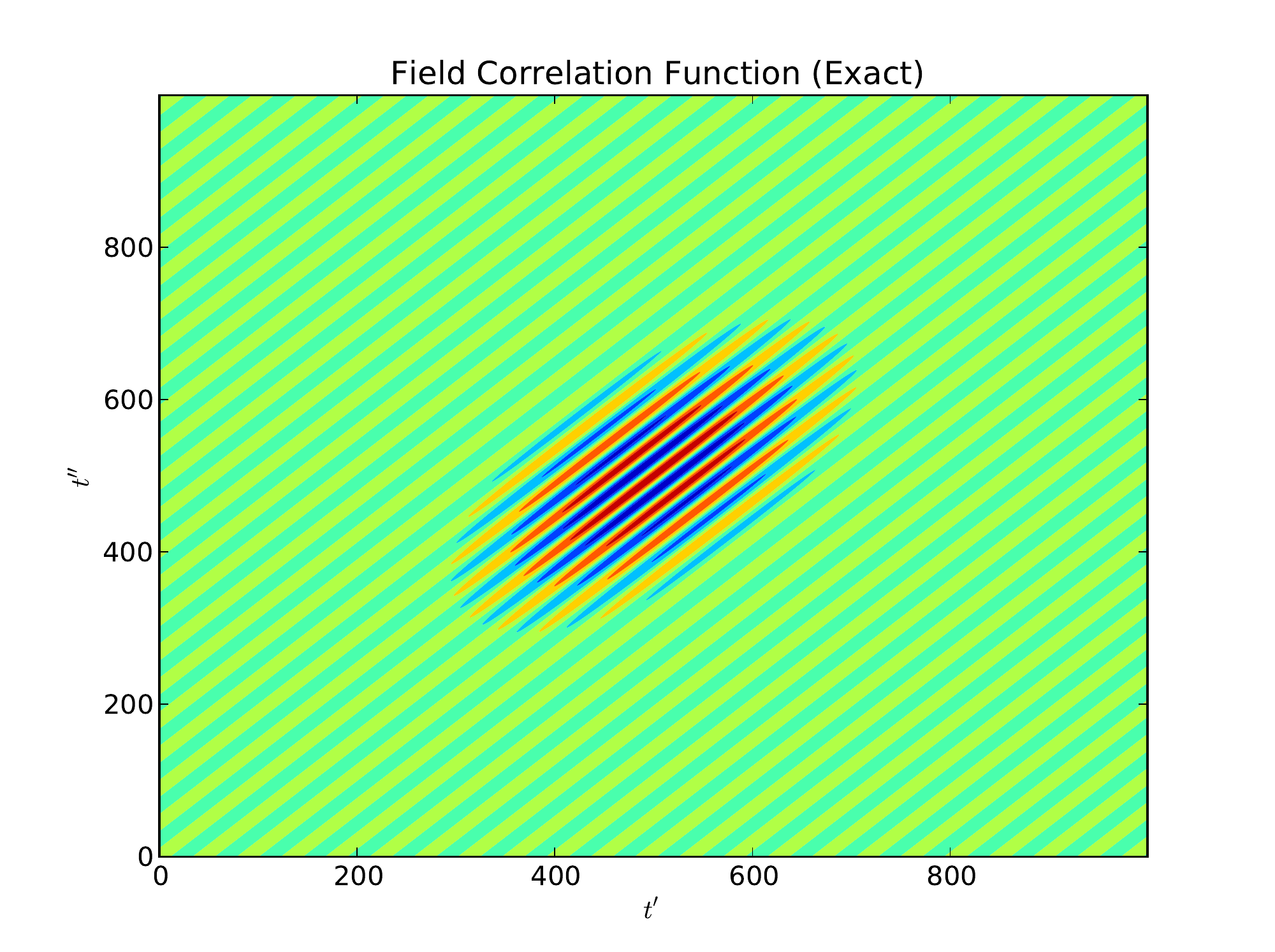}
\includegraphics[scale=0.5]{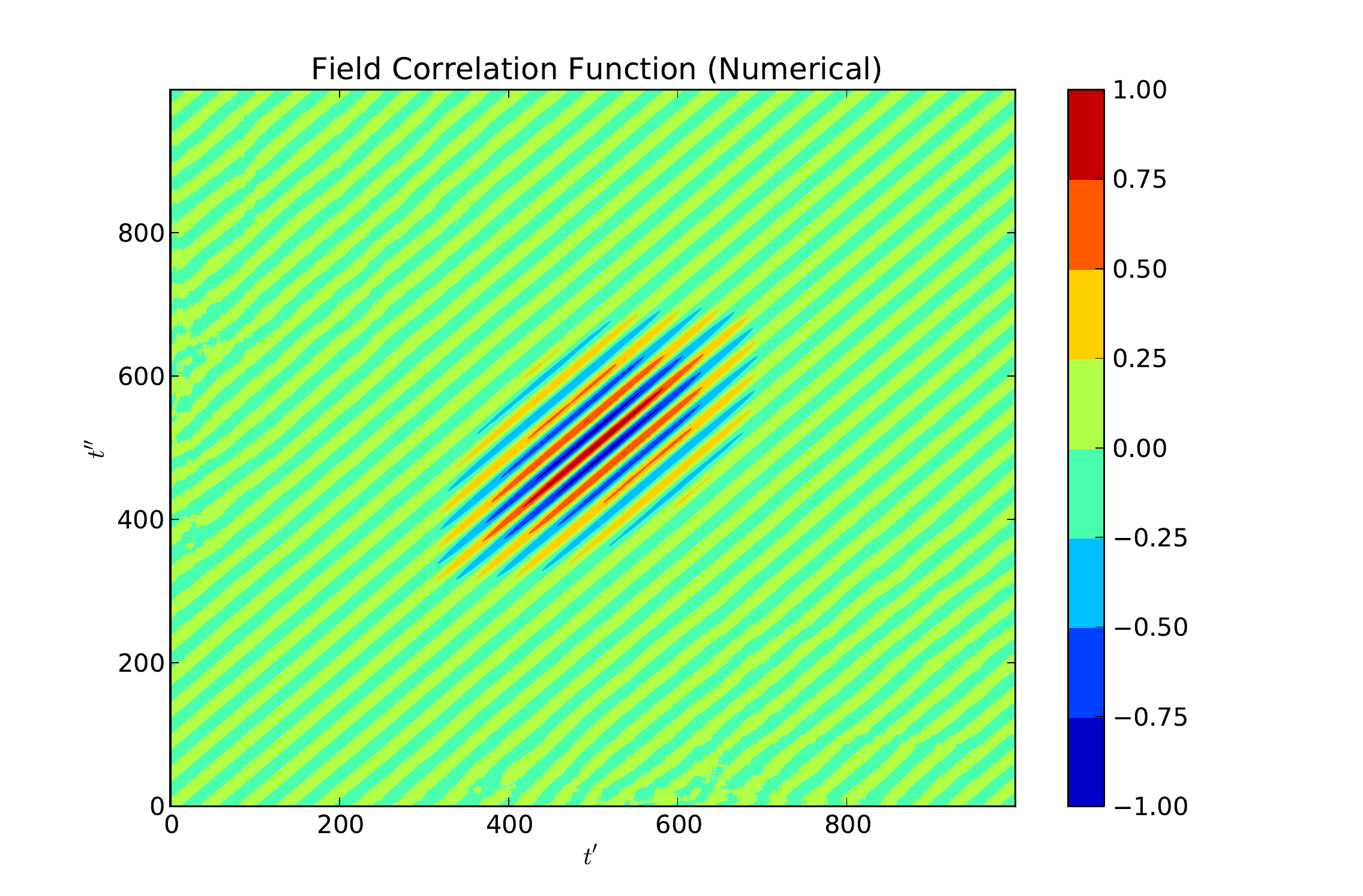}
\caption{(Top) Exact contour plot of the correlation function  in
Eq. \ref{corr2} and (Bottom) numerical reproduction of correlation
for the noisy pulsed source. Both plots share the same color
legend. The times are in units of femtosecond.} \label{NPexnum}
\end{center}
\end{figure}

\newpage
\begin{widetext}

\appendix

\section{Numerical Reproduction of Incoherent Light}
\subsection{Collisionally Broadened CW Laser}

The approach outlined in ref \cite{loudon,gardiner} is used to model the incoherent light. In this model the light has a central frequency $\omega_{0}$ and a time dependent phase $\phi(t)$ that abruptly changes $b$ times. The change in phase is taken from a uniform distribution.  The times at which these collisions occur, $\{ t_{j} \}$  are governed by a Wiener process $W(t)$ with a distribution center of zero and a drift coefficient ($D$) that is given by some scaling factor multiplied by the coherence time of the radiation $\tau_{d}$. From this process, a phase interrupted harmonic drive is generated. Each realization has $b$ collisions between the initial time $t_i$ to final time $t_f$. Several thousand of these realizations are generated, creating a set of 
electric fields $\{\varepsilon(t) \}$, from which the two time correlation function is computed. The real component of an ith realization is given by:

\begin{eqnarray}
&& \text{Re}(\varepsilon_{i}) (t) \propto \cos(\omega_{0}t + \phi_{i}(t)) \\
&& \phi_{i}(t) = \phi_{0}\theta(-t-t_{i}) + \prod_{k}{\theta(t-t_{k})\phi_{k}\theta(-t+t_{k+1})} + \phi_{f}\theta(t-t_{f})
\label{realization}
\end{eqnarray}
Here, $\theta(t)$ is the Heaviside theta function, $\phi_{m}$ are the phase changes and the set $\{ t_{j} \}$ are the phase interruption times.
For the case of $\tau_{d} = 120$ fs, $b$ is chosen to be a random number from a uniform distribution between 10 and 12. For different values of $\tau_{d}$ one has to adjust the drift coefficient $D$ and the number of collision events $b$ to reach agreement with the exact correlation function.

Results of this numerical method are plotted in Fig. \ref{CWexnum}
and compared to the exact expression [Eq. (\ref{corr1})]. It is
obvious that, although the numerical method can reproduce most of
the true correlation function, it still generates some error.
It should be noted that finding the correct parameters ($b$ and
$D$) to fit the exact correlation function numerically  was 
non-trivial and highly dependent on the radiation coherence time
$\tau_{d}$.

\subsection{Noisy Pulsed Light}

Generating noisy pulses is similar to that outlined above  except
that a Gaussian envelope is imposed on incoherent light.

\begin{eqnarray}
&& \text{Re}(\varepsilon_{i}) (t) \propto p(t)\cos(\omega_{0}t + \phi_{i}(t)) \\
&& p(t) = \left(\frac{2}{\pi \tau_{p}}\right)^{1/4} e^{-\frac{(t-t_{m})^2}{\tau_{p}}}
\end{eqnarray}
Here $\tau_{p}$ is the pulse duration and $t_{m}$ is the pulse center.
The results of this procedure  are shown in Fig. \ref{NPexnum} where the numerical and exact
correlation functions are compared. It is clear that noisy pulsed
source is sufficiently well reproduced by the numerical procedure.

\section{Transient Response as a Function of Turn-on Time for a Wiener CW Source}

The treatment in Eq. (\ref{den1}) assumes a sudden turn-on.  Of
interest is the effect of averaging over the turn-on, which may
well occur for an ensemble.  To this end, consider an arbitrary
turn-on time, $\tau_0$

\begin{eqnarray}
\rho_{e}&=&\sum_{i,j}c_{i}c_{j}^{*}|E_{i}\rangle\langle
E_{j}|e^{-i\omega_{ij}t}\int_{t_{0}}^{t}dt'e^{i\omega_{ig}t'}\int_{t_{0}}^{t}dt''e^{-i\omega_{jg}t''}\langle\varepsilon(t')\varepsilon^{*}(t'')\rangle
\\ & = &\sum_{i,j}c_{i}c_{j}^{*}|E_i\rangle\langle E_j|\Upsilon
\label{den2}
\end{eqnarray}

with  $\Upsilon = \Upsilon_{1} +
\Upsilon_{2} + \Upsilon_{3}$, where
\begin{eqnarray}
\Upsilon_{1}=\frac{2i\tau_{d}(1-e^{-i\omega_{ij}(t-t_{0})})}{\omega_{ji}(1+\tau_{d}^{2}\delta_{j}^{2})}
\end{eqnarray}

\begin{equation}
\Upsilon_{2}=\frac{\tau_{d}^{2}(1-e^{-\frac{(t-t_{0})}{\tau_{d}}}e^{-i\delta_{i}(t-t_{0})})}{(-i+\tau_{d}\delta_{j})(-i+\tau_{d}\delta_{i})}
\end{equation}

\begin{equation}
\Upsilon_{3}=\frac{\tau_{d}^{2}(e^{-i\omega_{ij}(t-t_{0})}-e^{-\frac{(t-t_{0})}{\tau_{d}}}e^{i\delta_{j}(t-t_{0})})}{(i+\tau_{d}\delta_{j})(i+\tau_{d}\delta_{i})}
\end{equation}

Consider then an ensemble average  over  $t_{0}$, where the
distribution of start times is assumed uniform, and the average is
taken over some time $[0,\frac{2\pi}{\omega_{ij}}]$. The terms
that are proportional to $e^{-i\omega_{ij}t_{0}}$ go to  zero
under this average while other terms are weighted by
$e^{-\frac{t}{\tau_{d}}}$ and decay on a time scale associated
with $\tau_{d}$. Averaging over these terms and neglecting terms
that decay for $t \gg\tau_{d}$ gives

\begin{equation}
\langle \rho_{e}^{ij} \rangle \propto \frac{2i\tau_{d}}{\omega_{ji}(1+\tau_{d}^{2}\delta_{j}^{2})}+\frac{\tau_{d}^{2}}{(-i+\tau_{d}\delta_{j})(-i+\tau_{d}\delta_{i})}
\label{statcoh}
\end{equation}

That is,  the
coherences are found to approach a nonzero value which is a
function of the level spacing, radiation coherence time  and
laser detuning.

\subsection{Maximizing Stationary Coherences}

Stationary coherences of the type seen in Eq. (\ref{statcoh}) are of interest in a number of contexts, e.g., one-photon phase control \cite{pachon-brumer}.  Maximizing this term is hence of interest.  For convenience, define  Eq.  (\ref{statcoh}) as $F$.
\begin{equation}
F = \frac{2i\tau_{d}}{\omega_{ji}(1+\tau_{d}^{2}\delta_{j}^{2})}+\frac{\tau_{d}^{2}}{(-i+\tau_{d}\delta_{j})(-i+\tau_{d}\delta_{i})}
\label{statcoh1}
\end{equation}
with
\begin{equation}
|F|^{2} = \frac{\tau_{d}^{2}\left(4+\omega_{ji}^{2}\tau_{d}^{2}\right)}{\omega_{ji}^{2}\left(1+\delta_{i}^{2}\tau_{d}^{2}\right)\left(1+\delta_{j}^{2}\tau_{d}^{2}\right)}
\label{absstatcoh}
\end{equation}
The extrema of Eq. (\ref{absstatcoh}) are then found with respect
to $\tau_{d}$ and $\omega_{0}$. The only physical extrema for
$\tau_{d}$ is $\tau_{d} = 0$ which minimizes coherences. With
respect to the laser frequency, the only physical extrema is
$\omega_{0} = (\omega_{ig}+\omega_{jg})/2$. In this case,
$|F|^{2}$ becomes:
\begin{equation}
|F|^{2} = \frac{16\tau_{d}^{2}}{\omega_{ji}^{2}\left(4+\tau_{d}^{2}\omega_{ji}^{2}\right)}
\end{equation}
Hence, as the excited state splitting decreases, this long-time coherence
becomes much larger.

As an example, using the parameters of PC645 (listed in Appendix D),  we plot
$|F|^{2}$ in Fig. \ref{absfsq}. As the coherence time is increased, it
is clear that this stationary coherence eventually saturates. In
the large coherence time and with $\omega_0 = (\omega_{ig} + \omega_{jg})/2$, this
stationary coherence becomes:
\begin{equation}
|F|^{2} = \frac{16}{\omega^{4}_{ji}}
\label{statcohinf}
\end{equation}

\section{White Noise Perturbative Result}

Most work on incoherent excitation has utilized  white noise
incident on model systems \cite{kozlov, scully}. White noise is a
mathematical construct with no real physical analog. We approach this
problem from the perturbative approach used in Eq. (\ref{den1}).
White noise is given by the following correlation function
\cite{kozlov}:
\begin{equation}
\langle\varepsilon(t')\varepsilon^{*}(t'')\rangle= \mathcal{R}\delta(t'-t'')
\label{whitenoise}
\end{equation}
The parameter $\mathcal{R}$ represents  the pump power of the
source. The Fourier transform of white noise contains all
frequencies uniformly,   exciting all states.

We use the same approach as outlined by Eq. (\ref{den1})  and
substitute Eq. (\ref{whitenoise}) for the correlation function of
the radiation field. Solving for the populations of the excited
states we get:
\begin{equation}
\rho_{e}^{ii}= \frac{|c_{i}|^{2}}{\hbar^{2}} \mathcal{R}t
\label{whitepop}
\end{equation}
Solving for coherences we get:
\begin{equation}
\rho_{e}^{ij}= \frac{c_{i}c^{\star}_{j}}{\hbar^{2}} \mathcal{R}\frac{1-e^{-i\omega_{ij}t}}{i\omega_{ij}}
\label{whitecoh}
\end{equation}

Linear population growth is seen similar to the growth  seen in
collisionally broadened CW case. Each excited state is pumped at
an equal rate (assuming that the dipole matrix elements of the
excited states are the same). The more intense the field the
larger the coherences between excited states. However, this comes
with the caveat that populations are also pumped at a faster rate.
Using these two equations (\ref{whitepop}-\ref{whitecoh}) the
mixed state measure, $\mathcal{C}$ for white noise irradiated $V$
level system is:
\begin{equation}
\mathcal{C} = \frac{|1-e^{-i\omega_{ij}t}|}{2\omega_{ij}t}
\end{equation}

This equation assumes that the transitions  between the ground
state to the excited states have the same dipole moment.
$\mathcal{C}$ is then plotted in Fig. \ref{whiteC}:

From Fig. \ref{whiteC}, it is clear that after several excited
state periods, the coherences of the system becomes a small
fraction of population, thus approaching a mixed state. It should
also be noted that this excited state coherence fraction is
independent of pump power, $\mathcal{R}$. This means that there
is no way to enhance the excited state coherence by simply
turning up the power of the source.

The collisionally broadened CW source (Eq. (\ref{corr1})) cannot
readily be related to white noise. It would be natural to assume that in
the limit $\tau_{d} \rightarrow 0$ the collisionally broadened CW
source should converge to white noise, but this is not the case.
In that limit, $\mathcal{U} \rightarrow 0$ and thus the frequency
spectrum of the laser also approaches zero. White noise, on the
other hand, has a uniform frequency spectrum.

\section{Parameters Used in PC645 Calculation}

\begin{center}
\begin{table}
    \begin{tabular}{ | l | p{5cm} |}
    \hline
    Parameter & Value \\ \hline
    $DBV^{+}$ Frequency & 529 THz \\ \hline
    $DBV^{-}$ Frequency & 510 THz\\ \hline
    $\tau_{d}$ & 1.32 fs\\ \hline
    $\omega_{0}$ Frequency & 519.5 THz \\ \hline
    $\mu$  & 12.8 Debye \\ \hline
    Solar Flux &  130,000 lux\\ \hline
    \end{tabular}
\caption{Parameters used in perturbative calculations of PC645.}
\label{tabofpc645}
\end{table}
\end{center}

The parameters used in the perturbative calculation of the
incoherent irradiation of PC645 are listed in Table
\ref{tabofpc645}. The values for the $DBV^{\pm}$ states as well as
dipole moment were taken from the literature \cite{sch1}. The
dipole moment of the two DBV c,d sites were averaged and used for
as the dipole moment of the  $DBV^{\pm}$ states. This is valid as
long as there is no dipole matrix element between the DBVc,d
states, i.e. $\langle DBVc| \mu | DBVd \rangle = 0$. Values of the
coherence time of the radiation was taken from the approach of
Wolf \cite{wolf} and the solar flux was taken from values of the
solar flux in midday \cite{sunvalues}.

\end{widetext}

\newpage
\section{Figures}

 \begin{figure}[h]
\begin{center}
\includegraphics[scale=0.5]{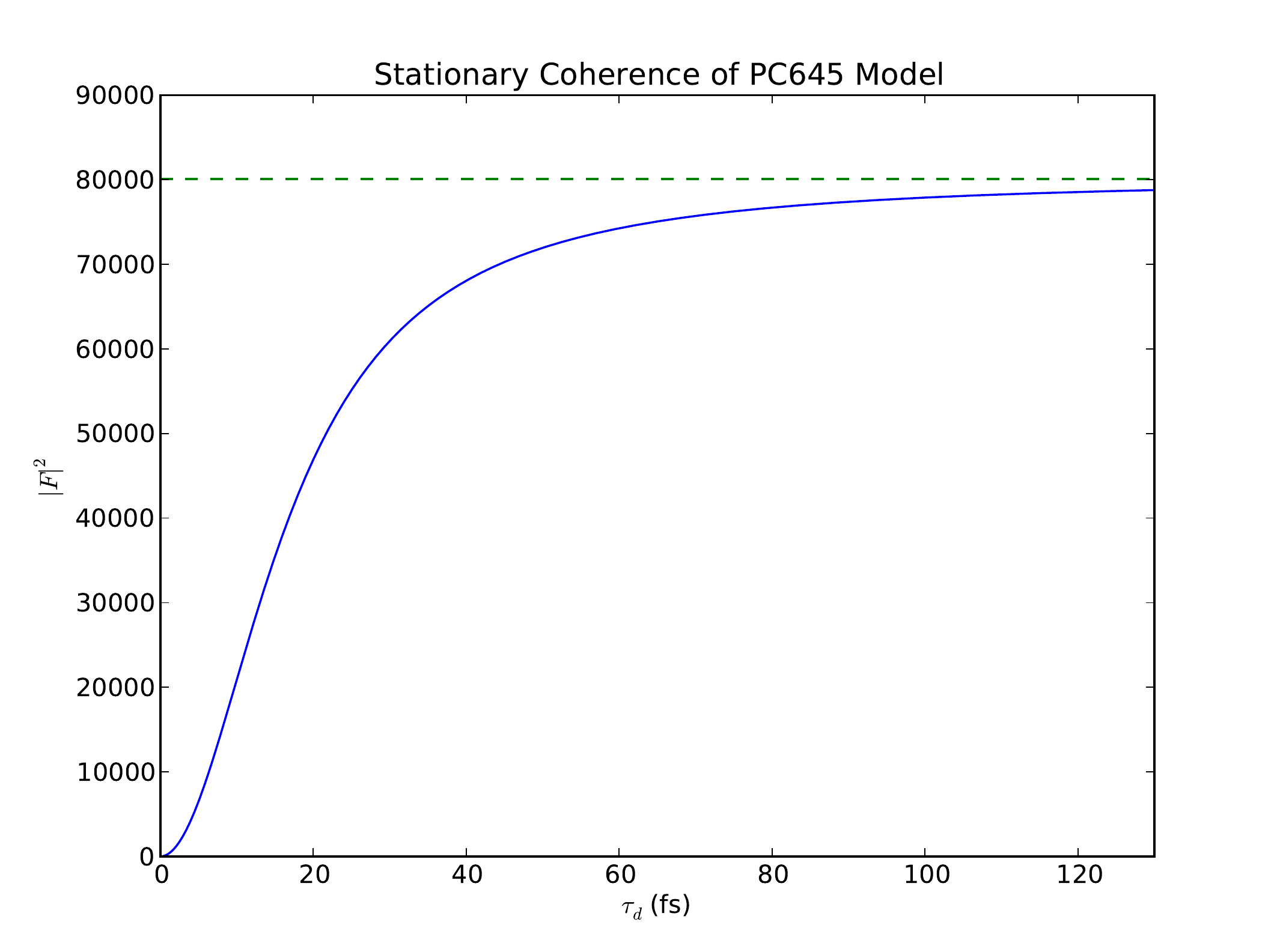}
\caption{Plot of $|F|^{2}$ using the parameters of PC645.  Solid
line is the stationary coherence as a function of $\tau_{d}$ while
the dashed line represents the saturation level from Eq.
(\ref{statcohinf}).} \label{absfsq}
\end{center}
\end{figure}

 \begin{figure}[h]
\begin{center}
\includegraphics[scale=0.5]{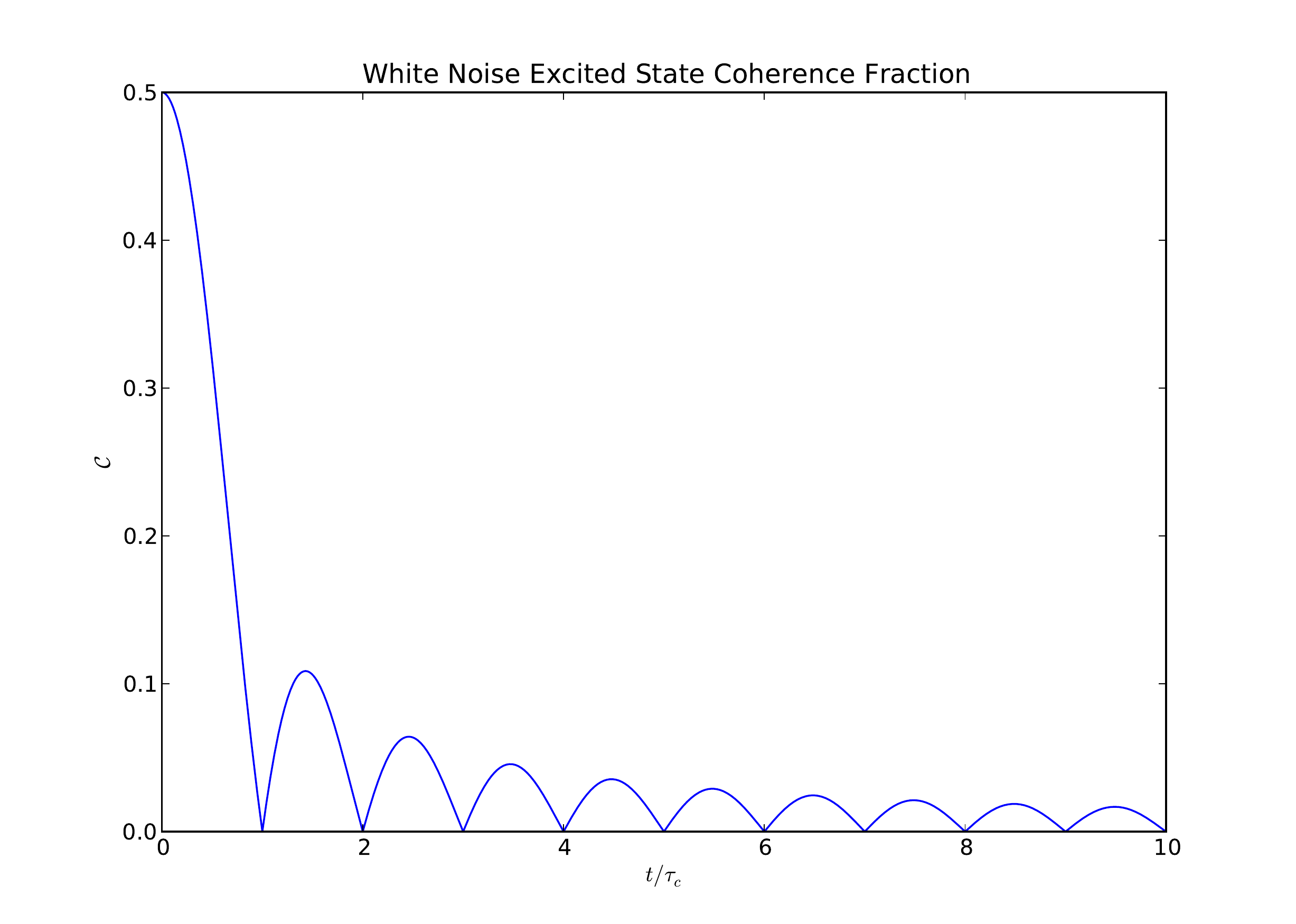}
\caption{Plot of $\mathcal{C}=\frac{|\rho_{23}|}{\rho_{33} + \rho_{22}}$ for a three level system irradiated with white noise as a function of dimensionless time. $\tau_{c} = \frac{2\pi}{\omega_{ij}} $}
\label{whiteC}
\end{center}
\end{figure}


\newpage
\begin{thebibliography}{100}
\bibitem{col} E. Collini, C. Y. Wong, K. E.Wilk, P. M. G. Curmi, P. Brumer and G. D. Scholes, Nature {\bf 463}, 644 (2010).
\bibitem{engel} G. S. Engel, T. R. Calhoun, E. L. Read, T.-K. Ahn, T. Mancal, Y.-C. Cheng, R. E. Blankenship and G. R. Fleming, Nature {\bf 446}, 782 (2007).
\bibitem{brumer} X.-P. Jiang and P. Brumer, J. Chem. Phys. {\bf 94}, 5833 (1991).
\bibitem{onephoton} P. Brumer and M. Shapiro,  Proc. Natl. Acad. Sci. USA \textbf{109}, 19575 (2012)
\bibitem{mancal} T. Mancal and L. Valkunas, New J. Phys. {\bf 12}, 065044 (2010).
\bibitem{leo} L. A. Pachon and P. Brumer, Phys. Rev. A  \textbf{87}, 022106 (2013).
\bibitem{Han} A. Han, M. Shapiro and P. Brumer, J. Phys. Chem. A \textbf{117}, 8199 (2013). 
\bibitem{aharony} A. Aharony, S. Gurvitz, O. Entin-Wohlman and S. Dattagupta, Phys. Rev. B {\bf 82}, 245417, (2010).
\bibitem{kozlov} V. V. Kozlov, Y. Rostovtsev and M. O. Scully, Phys. Rev. A {\bf 74}, 063829 (2006).
\bibitem{scully} M. Fleischhauer, C. H. Keitel and M. O. Scully, Opt. Commun. {\bf 87}, 109 (1992).
\bibitem{Elran1} Y. Elran and P. Brumer, J. Chem. Phys. \textbf{121}, 2673 (2004)
\bibitem{Elran2} Y. Elran and P, Brumer, J. Chem. Phys. \textbf{138}, 234308 (2013)
\bibitem{loudon} R. Loudon, {\it The Quantum Theory of Light}, Oxford University Press, Oxford (1983).
\bibitem{wolf} C. L. Mehta and E. Wolf, Phys. Rev. {\bf 134}, A1143 (1964).
\bibitem{mandelwolf} L. Mandel and E. Wolf, {\it Optical Coherence and Quantum Optics}, Cambridge University Press, New York (1995).
\bibitem{brumerbook} M. Shapiro, P. Brumer, {\it Principles of the Quantum Control of Molecular Processes},  Wiley, New York (2003); M. Shapiro and P. Brumer, {\it Quantum Control of Molecular Processes}, 2nd edn, Wiley-VCH, Weinheim (2012).
\bibitem{gardiner} C.W. Gardiner, {\it Handbook of Stochastic Methods for Physics, Chemistry and Natural Sciences}, Springer Verlag, Berlin (1983). Chapter IV, page 257.
\bibitem{fleming} Y-C Cheng, G.R. Fleming, Annu. Rev. Phys. Chem. {\bf 60}, 241 (2009).
\bibitem{pachon-brumer} L. A. Pachon, L. Yu and P. Brumer, Faraday Disc. {\bf 163},
485 (2013); L. A. Pachon and P. Brumer, J. Chem. {\bf 139}, 164123 (2013).
\bibitem{scho} G. D. Scholes, G. R. Fleming, A. Olaya-Castro and R. van Grondelle, Nature {\bf 3}, 763 (2011).
\bibitem{sch1} A. B. Doust, K. E. Wilk, P. M. G. Curmi and G. D. Scholes, J. Photochem.  Photobio. A {\bf 184}, 1 (2006).
\bibitem{sunvalues} P. Schlyter,  \textit{Radiometry and Photometry in Astronomy}, http://stjarnhimlen.se/comp/radfaq.html Stockholm  (2009).
\bibitem{grinev} T. Grinev and P. Brumer (manuscript in preparation)
\end{thebibliography}
\end{document}